\documentclass[11pt]{article}

\usepackage[margin=1in]{geometry}
\usepackage{amsmath,amssymb,amsthm,mathtools}
\usepackage{enumitem}
\usepackage{listings}
\lstdefinestyle{pythonappendix}{
  language=Python,
  basicstyle=\footnotesize\ttfamily,
  keywordstyle=\bfseries,
  commentstyle=\itshape,
  stringstyle=\ttfamily,
  showstringspaces=false,
  columns=fullflexible,
  keepspaces=true,
  breaklines=true,
  breakatwhitespace=false,
  numbers=left,
  numberstyle=\tiny,
  stepnumber=5,
  numbersep=8pt,
  xleftmargin=20pt,
  frame=lines,
  aboveskip=\medskipamount,
  belowskip=\medskipamount,
  tabsize=4
}
\usepackage[textsize=footnotesize]{todonotes}
\usepackage[hidelinks]{hyperref}
\usepackage[nameinlink,capitalize]{cleveref}

\newtheorem{theorem}{Theorem}[section]

\newtheorem{lemma}[theorem]{Lemma}

\newtheorem{corollary}[theorem]{Corollary}

\newtheorem{proposition}[theorem]{Proposition}

\newtheorem{remark}[theorem]{Remark}

\newtheorem{definition}[theorem]{Definition}

\newcommand{\F}{\mathbb{F}}

\newcommand{\BS}{\operatorname{BS}}

\newcommand{\BU}{\operatorname{BU}}

\newcommand{\disc}{\operatorname{disc}}

\crefname{lemma}{Lemma}{Lemmas}
\Crefname{lemma}{Lemma}{Lemmas}
\crefname{proposition}{Proposition}{Propositions}
\Crefname{proposition}{Proposition}{Propositions}
\crefname{corollary}{Corollary}{Corollaries}
\Crefname{corollary}{Corollary}{Corollaries}
\crefname{theorem}{Theorem}{Theorems}
\Crefname{theorem}{Theorem}{Theorems}
\crefname{definition}{Definition}{Definitions}
\Crefname{definition}{Definition}{Definitions}

\title{\huge\bfseries On the boomerang spectrum of the APN power function $x^{q+2}$ over $\F_{q^2}$}
\author{\Large\bfseries Kirpa Garg$^1$, Constanza Riera$^2$ and Pantelimon St\u{a}nic\u{a}$^3$\footnote{Pantelimon St\u{a}nic\u{a} is the corresponding author.}\\
\vspace{.2cm}
\\[4pt]
$^1$University of Rouen Normandy - LITIS UR 4108,\\
Avenue de l’Université 76800 Saint-Étienne-du-Rouvray, \\
France,
\texttt{kirpa.garg@univ-rouen.fr}\\[4pt]
$^2$Department of Computer Science,\\
Electrical Engineering and Mathematical Sciences,\\
Western Norway University of Applied Sciences,\\
5020 Bergen, Norway,
\texttt{csr@hvl.no}\\[4pt]
$^3$Applied Mathematics Department, Naval Postgraduate School,\\
Monterey, CA 93943, USA,
\texttt{pstanica@nps.edu}}
\date{\today}

\begin{document}
\maketitle

\begin{abstract}
We study the boomerang spectrum of the almost perfect nonlinear power function $x^{q+2}$ over $\F_{q^2}$, where $q$ is an odd prime power with $q\equiv1\pmod6$. The connection of Pal and St\u{a}nic\u{a} between boomerang entries and $(-1)$-differential entries reduces the problem to counting roots of a single equation. We determine the complete contribution from nonzero parameters in $\F_q$. For parameters outside $\F_q$, we obtain a quintic over $\F_q$ in terms of the parameter's trace and norm, with a separate formula when the trace is zero. We exclude the entry four when $q\equiv1\pmod{12}$ and characterize its occurrence when $q\equiv7\pmod{12}$ through a repeated-root parametrization and a residual cubic. We also evaluate the first nonsubfield moment and an aggregate quadratic-character sum, and prove a Weil bound for each nonzero trace. Explicit secant and residual-cubic formulas express the second, third, and fourth falling-factorial moments as finite sums. Once the size-four frequency is known, these moments recover all remaining frequencies through a nonsingular five-dimensional system. A closed-form evaluation of the nonsubfield distribution remains open.
\end{abstract}



\section{Introduction}
\label{sec:intro}

Throughout this paper, $q=p^m$ is an odd prime power with $q\equiv1\pmod6$. We write $\F_q^*=\F_q\setminus\{0\}$ and use $\#$ for cardinality. The quadratic character $\eta$ of $\F_q$ is extended by $\eta(0)=0$. Thus $\eta$ takes the values $1$ and $-1$ on the sets $C_0$ and $C_1$ of nonzero squares and nonsquares, respectively.

The boomerang attack, introduced by Wagner~\cite{Wagner1999}, combines related differential trails through a block cipher. Cid et al.~\cite{CidEtAl2018} introduced the boomerang connectivity table, and Boura and Canteaut~\cite{BouraCanteaut2018} studied the associated boomerang uniformity in the binary setting. We use the coupled-system formulation~\cite{LiQuSunLi2019} and its extension to odd characteristic and to functions that need not be permutations~\cite{PalStanica2025}. For a function $f\colon\F_Q\to\F_Q$, where $Q$ is an odd prime power, let $\beta_f(a,b)$ be the number of pairs $(x,y)\in\F_Q^2$ satisfying
\begin{equation}
\label{eq:boomerang-system}
  f(x)-f(y)=b,
  \qquad
  f(x+a)-f(y+a)=b.
\end{equation}
The boomerang uniformity is the maximum of these counts over $a,b\in\F_Q^*$. The boomerang spectrum records their frequencies. It therefore distinguishes functions whose largest entries agree but whose entry distributions differ.

We consider $F(x)=x^{q+2}$ over $\F_{q^2}$. Under our hypothesis, $F$ is almost perfect nonlinear (APN), meaning that every equation $F(x+a)-F(x)=b$ with $a\ne0$ has at most two solutions~\cite[Section~2]{PalStanica2025}. Mesnager and Wu~\cite{MesnagerWu2025} determined its boomerang uniformity for $q\equiv1$ and $q\equiv3\pmod6$, and obtained further spectrum information when $q\equiv3\pmod6$. Here we study the distribution for $q\equiv1\pmod6$. We determine the subfield contribution and reduce the nonsubfield contribution to explicit root counts and moment sums.

The power-map scaling symmetry reduces the boomerang entries to $a=1$. Since $F$ is odd and APN, the connection of Pal and St\u{a}nic\u{a}~\cite{PalStanica2025} then gives
\begin{equation}
\label{eq:E-b}
  \beta_F(1,b)=\#\{X\in\F_{q^2}:4X^{q+2}+X^q+2X+2b=0\}.
\end{equation}
For $b\in\F_q^*$, conjugation separates this equation into a cubic over $\F_q$ and a quadratic norm branch. Stickelberger's parity theorem and an explicit cube-class count give the complete distribution of their combined root counts.

For $b\in\F_{q^2}\setminus\F_q$, put $A=-(b+b^q)$ and $B=b^{q+1}$, the negative trace and norm of $b$. The two conjugate parameters $b$ and $b^q$ have the same root count, denoted by $N(A,B)$. Their coefficient pairs form the set $\mathcal D'=\{(A,B)\in\F_q^2:A^2-4B\in C_1\}$. Writing $X=u+vb$ with $u,v\in\F_q$ gives a quintic in $v$ when $A\ne0$, while $A=0$ requires a separate cubic and quadratic count. These reductions bound every entry by five. A repeated-root calculation excludes $N(A,B)=4$ when $q\equiv1\pmod{12}$. When $q\equiv7\pmod{12}$, it characterizes the nonzero-trace pairs with $N(A,B)=4$ through a rational parametrization and the complete splitting of a residual cubic. In the trace-zero case, the value four occurs precisely when $B=-1/108$ and $\eta(5)=1$.

For $r\ge0$, let $M_r=\sum_{(A,B)\in\mathcal D'}N(A,B)^r$, with each summand interpreted as $1$ when $r=0$. We evaluate $M_1$ exactly and derive both a fixed-trace character-sum bound and an exact aggregate character sum. Secant interpolation expresses the higher falling-factorial moments as finite sums, with a residual cubic describing the roots that remain after two are fixed. Together with the size-four frequency, the moments through order four determine the other frequencies through a nonsingular $5\times5$ system. The evaluation of these higher sums, and hence a closed-form nonsubfield distribution, remains open.

Section~\ref{sec:definitions} gives the definitions and auxiliary results, and Section~\ref{sec:first-reduction} derives the single equation in~\eqref{eq:E-b}. Section~\ref{sec:subfield} determines the subfield contribution. Section~\ref{sec:nonsubfield} develops the trace--norm reduction and the size-four characterization, and Section~\ref{sec:incidence-bound} proves the fixed-trace incidence bound. Section~\ref{sec:geometry} describes the rational curves and derives the moment formulas. Section~\ref{sec:nonsubfield-consequences} converts the coefficient-pair counts into the boomerang spectrum, and Section~\ref{sec:conclusion} identifies the remaining counting problems.

\section{Definitions and auxiliary tools}
\label{sec:definitions}

Let $Q$ be an odd prime power. For a function $f:\F_Q\to\F_Q$, its differential uniformity is $\max_{a\in\F_Q^*,\,b\in\F_Q}\#\{x:f(x+a)-f(x)=b\}$. A function of differential uniformity two is called almost perfect nonlinear (APN). A function is odd if $f(-x)=-f(x)$ for every $x$. Here $\F_Q^*=\F_Q\setminus\{0\}$ and $\#$ denotes cardinality.

Let $f:\F_Q\to\F_Q$ be any function.  For $a,b\in\F_Q^*$ define
\[
  \beta_f(a,b)=\#\{(x,y)\in\F_Q^2: f(x)-f(y)=b,
  \ f(x+a)-f(y+a)=b\}.
\]
The boomerang uniformity and boomerang spectrum are
$\BU(f)=\max_{a,b\in\F_Q^*}\beta_f(a,b), \ \BS(f)=(\nu_i)_{0\le i\le \BU(f)},$
where
$\nu_i=\#\{(a,b)\in\F_Q^*\times\F_Q^*: \beta_f(a,b)=i\}.$
For a power function we use the following standard scaling reduction.

\begin{lemma}
\label{lem:power-normalization}
Let $f(x)=x^d$ over $\F_Q$.  Then, for every $a,b\in\F_Q^*$,
$\beta_f(a,b)=\beta_f\left(1,\frac{b}{a^d}\right).$
Consequently, if
$n_i=\#\{c\in\F_Q^*: \beta_f(1,c)=i\},$
then $\nu_i=(Q-1)n_i$.
\end{lemma}

\begin{proof}
Put $x=az$ and $y=aw$.  Then
$f(x)-f(y)=a^d(z^d-w^d),  f(x+a)-f(y+a)=a^d((z+1)^d-(w+1)^d).$
Thus the system defining $\beta_f(a,b)$ is carried bijectively to the system defining $\beta_f(1,b/a^d)$.
\end{proof}

We also use the elementary quadratic-character identity~\cite[Theorem 5.48]{LidlNiederreiter}
\begin{equation}
\label{eq:quadratic-character-identity}
\sum_{t\in\F_q}\eta(t^2-c)=-1\qquad(c\in\F_q^*),
\end{equation}
which follows by counting factorizations. Indeed, the number of pairs $(t,y)$ satisfying $y^2=t^2-c$ is $q+\sum_t\eta(t^2-c)$. The bijection $(t,y)\mapsto(t-y,t+y)$ identifies these pairs with the $q-1$ factorizations of $c$ into two nonzero factors, proving the identity. Completing the square gives the corresponding identity for every quadratic with nonzero discriminant.

Under $q\equiv1\pmod6$, choose a primitive cube root of unity $\omega\in\F_q$. Since $(2\omega+1)^2=-3$, one has $\eta(-3)=1$ and $\eta(3)=\eta(-1)$.

We shall use the following form of the Weil bound for multiplicative character sums applied to the quadratic character $\eta$ (see \cite[Theorem~5.41]{LidlNiederreiter}).
The condition stated below that $P$ is not of the form $cH^2$ means simply that
  $P$ is not a perfect square (up to a nonzero scalar), since $\eta$ has order~$2$.
Multiplying $P$ by a nonzero scalar changes the sum only by the constant factor $\eta(c)$, and hence does not change its absolute value.

\begin{theorem}
\label{thm:weil-polynomial-character}
Let $P\in\F_q[V]$ be a polynomial of positive degree which is not of the form $cH(V)^2$ with $c\in\F_q^*$ and $H\in\F_q[V]$.  Let $r$ be the number of distinct roots of $P$ in an algebraic closure of $\F_q$.  Then
$\left|\sum_{v\in\F_q}\eta(P(v))\right|\le (r-1)\sqrt q.$
In particular, if $\deg P=d$, then
$\left|\sum_{v\in\F_q}\eta(P(v))\right|\le (d-1)\sqrt q.$
\end{theorem}

We use the $c$-Difference Distribution Table and the Pal--St\u{a}nic\u{a} connection to reduce the boomerang spectrum to explicit algebraic counting problems.

\begin{definition}[\cite{EFRST20}]
Let $F$ be a function from $\F_{p^n}$ to itself and let $c\in\F_{p^n}$. The $c$-Difference Distribution Table ($c$-DDT) entries at $(a,b)\in\F_{p^n}\times\F_{p^n}$ are defined by ${}_{c}\Delta_F(a,b)=\#\{x\in\F_{p^n}:F(x+a)-cF(x)=b\}$.
\end{definition}

The following result is the Pal--St\u{a}nic\u{a} connection~\cite{PalStanica2025}.

\begin{theorem}
\label{bridge}
For any odd APN function $F$ over $\F_{p^n}$, where $p$ is an odd prime,
$\beta_F(a,b)={}_{-1}\Delta_F(a,-b)$ for all $a,b\in\F_{p^n}^*$.
\end{theorem}

\section{First reduction}
\label{sec:first-reduction}

We first convert the boomerang entries into a single one-parameter equation over $\F_{q^2}$.  Since $F(x)=x^{q+2}$ is an odd APN power map, its boomerang entries are $(-1)$-differential entries.  After centering the variable, this gives the equation used throughout the rest of the paper.

\begin{proposition}
\label{prop:working-reduction}
Let $q=p^m$ be odd with $q\equiv1\pmod3$, and let $F(x)=x^{q+2}$ over $\F_{q^2}$.  Then $F$ is an odd APN function.  For every $b\in\F_{q^2}^*$,
\begin{equation}
\label{eq:centered-bridge}
  \beta_F(1,b)
  =\#\left\{X\in\F_{q^2}:\left(X+\frac12\right)^{q+2}+\left(X-\frac12\right)^{q+2}=-b\right\}.
\end{equation}
Equivalently, the nontrivial boomerang spectrum of $F$ is determined by the numbers of solutions in $\F_{q^2}$ of
\begin{equation}
\label{eq:main-equation}
  E_b(X):=4X^{q+2}+X^q+2X+2b=0,
  \qquad b\in\F_{q^2}^*.
\end{equation}
Let $E_b(\F_{q^2})$ denote the set of roots of \eqref{eq:main-equation}. If $n_i=\#\{b\in\F_{q^2}^*: \#E_b(\F_{q^2})=i\}$,
then the boomerang spectrum entries are  $\nu_i=(q^2-1)n_i$.
\end{proposition}

\begin{proof}
The exponent $q+2$ is odd, because $q$ is odd.  Hence
$F(-x)=(-x)^{q+2}=-x^{q+2}=-F(x),$
so $F$ is an odd function.  The APN property is the odd-characteristic family $x^{p^{n/2}+2}$ with $n=2m$ and $p^{n/2}=q\equiv1\pmod3$. This is the family denoted $f_4$ in the list of odd APN power maps used by Pal and St\u{a}nic\u{a} \cite[Section~2]{PalStanica2025}.
Thus, by Theorem \ref{bridge}, for an odd APN function over a finite field of odd characteristic one has
$\beta_F(a,b)={}_{-1}\Delta_F(a,-b) \ (a,b\in\F_{q^2}^*).$
Taking $a=1$ and using the centered variable gives
${}_{-1}\Delta_F(1,-b) =\#\left\{X\in\F_{q^2}:F\left(X+\frac12\right)+F\left(X-\frac12\right)=-b\right\},$
which proves \eqref{eq:centered-bridge}.  Expanding the left-hand side gives
$\left(X+\frac12\right)^{q+2}+\left(X-\frac12\right)^{q+2} =2X^{q+2}+\frac12X^q+X,$
because $q$ is odd and $(1/2)^q=1/2$ in $\F_q$.  Multiplication of this equation by $2$ gives \eqref{eq:main-equation}.  The formula for $\nu_i$ follows from the power-map normalization in \Cref{lem:power-normalization}, because for fixed $a \ne0$, the map $b\mapsto b/a^{q+2}$ permutes $\F_{q^2}^*$.
\end{proof}

We count the roots first for $b\in\F_q^*$ and then for $b\in\F_{q^2}\setminus\F_q$.

\section{The subfield contribution}
\label{sec:subfield}

In this section, let $b\in\F_q$.
When $b\in\F_q$, raising \eqref{eq:main-equation} to the $q$th power (using $X^{q^2}=X$ and $b^q=b$) and subtracting the two equations gives
\begin{equation}
\label{eq:factor-subfield}
  (4X^{q+1}+1)(X^q-X)=0.
\end{equation}

Thus every solution either lies in $\F_q$ or satisfies $X^{q+1}=-1/4$. We count the two branches and check their overlap. If $X\in\F_q$, equation~\eqref{eq:main-equation} reduces to $X^3+\frac34X+\frac b2=0$.
Let
\begin{equation}
\label{eq:G-b}
  G_b(X)=X^3+\frac34X+\frac b2.
\end{equation}
The discriminant is
\begin{equation}
\label{eq:disc-Gb}
  \disc(G_b)=-\frac{27}{16}(1+4b^2).
\end{equation}
Since $q\equiv1\pmod3$, $-3$ is a square in $\F_q$.  Therefore the square/nonsquare status of \eqref{eq:disc-Gb} is the same as that of $1+4b^2$.  For the rest of the subfield analysis we write
$\delta_b=1+4b^2.$ When $b=0$, $\delta_b=1$, so $\eta(\delta_b)=1$. We will study the case $b\in\F_q^*$ in the following lemma.

\begin{lemma}
\label{lem:N1N2}
Let $q\equiv1\pmod6$.  The number of $b\in\F_q^*$ for which $\delta_b=1+4b^2$ is a nonzero square in $\F_q$ is
\[
  N_1=
  \begin{cases}
    (q-5)/2,& q\equiv1\pmod{12},\\
    (q-3)/2,& q\equiv7\pmod{12}.
  \end{cases}
\]
The number of $b\in\F_q^*$ for which $\delta_b=1+4b^2$ is a nonsquare in $\F_q$ is
\[
  N_2=
  \begin{cases}
    (q-1)/2,& q\equiv1\pmod{12},\\
    (q+1)/2,& q\equiv7\pmod{12}.
  \end{cases}
\]
\end{lemma}

\begin{proof}
Let $t=2b$.  Since multiplication by $2$ is a bijection of $\F_q^*$, the required number is
$N_1=\#\{t\in\F_q^*: \eta(1+t^2)=1\}.$
Let
$Z=\#\{t\in\F_q^*:1+t^2=0\}.$
Then
\[
  N_1=\frac12\sum_{t\in\F_q^*}\bigl(1+\eta(1+t^2)\bigr)-\frac Z2
      =\frac{q-1-Z}{2}+\frac12\sum_{t\in\F_q^*}\eta(1+t^2).
\]
The polynomial $t^2+1$ has leading coefficient $1$ and discriminant $-4 \ne0$.  Identity~\eqref{eq:quadratic-character-identity} with $c=-1$ gives
$\sum_{t\in\F_q}\eta(t^2+1)=-\eta(1)=-1.$
The term $t=0$ contributes $\eta(1)=1$, hence
$\sum_{t\in\F_q^*}\eta(1+t^2)=-2.$
Therefore
$N_1=\frac{q-1-Z}{2}-1=\frac{q-3-Z}{2}.$
Now $Z=2$ exactly when $-1$ is a square in $\F_q$, and $Z=0$ otherwise.  Since $q\equiv1\pmod6$, this means $Z=2$ for $q\equiv1\pmod{12}$ and $Z=0$ for $q\equiv7\pmod{12}$.  This proves the displayed values of $N_1$, which counts exactly the nonzero-square cases.

For $N_2$, the square, nonsquare, and zero cases partition $\F_q^*$, and hence
$N_2=q-1-Z-N_1.$
Substituting the value of $Z$ gives
$N_2=\frac{q-1}{2}\quad(q\equiv1\pmod{12}),   N_2=\frac{q+1}{2}\quad(q\equiv7\pmod{12}).$
The proof is complete.
\end{proof}

Among $b\in\F_q^*$ with $\delta_b=1+4b^2$ a nonzero square, the number giving three roots of the cubic is $(q-13)/6$ when $q\equiv1\pmod{12}$ and $(q-7)/6$ when $q\equiv7\pmod{12}$.  We give a self-contained proof using the following results on cubics.

Stickelberger's parity theorem will be used in the following form~\cite{Stick}.

\begin{theorem}\label{ST}
Let $f\in\F_q[X]$ be squarefree of degree $d$, where $q$ is odd, and let $k$ be the number of its irreducible factors over $\F_q$. Then $\disc(f)$ is a square in $\F_q^*$ if and only if $k\equiv d\pmod2$.
\end{theorem}

\begin{proof}
Let $a$ be the leading coefficient and let $\alpha_1,\ldots,\alpha_d$ be the distinct roots in an algebraic closure. The element $\Delta=a^{d-1}\prod_{i<j}(\alpha_i-\alpha_j)$ satisfies $\Delta^2=\disc(f)$. Frobenius permutes the roots in $k$ cycles, one for each irreducible factor, so $\Delta^q=(-1)^{d-k}\Delta$. Thus $\Delta\in\F_q$ exactly when $d-k$ is even. Since $\Delta\ne0$, this is equivalent to its square being a square in $\F_q$.
\end{proof}

\begin{theorem}
\label{thm:cubic-count}
Let $q\equiv1\pmod6$.  Among the elements $b\in\F_q^*$ for which $1+4b^2$ is a nonzero square, the number for which the cubic $G_b(X)$ in \eqref{eq:G-b} splits completely over $\F_q$ is
\[
  C_q=
  \begin{cases}
    (q-13)/6,& q\equiv1\pmod{12},\\
    (q-7)/6,& q\equiv7\pmod{12}.
  \end{cases}
\]
The remaining such $b$ give no root of $G_b$ in $\F_q$.
\end{theorem}

\begin{proof}
Assume first that $\theta_b:=b^2+1/4$ is a nonzero square in $\F_q$, and choose $s\in\F_q^*$ with $s^2=\theta_b$.  Set $\alpha=(-b+s)/4$ and $\gamma=(-b-s)/4$. Then $\alpha+\gamma=-b/2$ and $\alpha\gamma=-1/64$. We first prove that $G_b$ splits over $\F_q$ if and only if $\alpha$ is a cube in $\F_q^*$.

If $a^3=\alpha$ with $a\in\F_q^*$, put $c=-1/(4a)$. Then $c^3=\gamma$, and $(a+c)^3+(3/4)(a+c)+b/2=0$. Since $q\equiv1\pmod3$, choose a primitive cube root of unity $\omega\in\F_q$. Replacing $(a,c)$ by $(\omega^j a,\omega^{-j}c)$, $j=0,1,2$, gives all three roots. They are distinct because $\disc(G_b)\ne0$.

Conversely, if $x\in\F_q$ is a root, choose roots $a,c$ of $T^2-xT-1/4$ in an algebraic closure. Then $a+c=x$ and $ac=-1/4$, so $a^3+c^3=-b/2$ and $a^3c^3=-1/64$. Hence $\{a^3,c^3\}=\{\alpha,\gamma\}$. If $a\notin\F_q$, Frobenius interchanges $a$ and $c$, and consequently $a^3=c^3$ because both cubes lie in $\F_q$. This contradicts $s\ne0$. Thus $a,c\in\F_q$, proving that $\alpha$ is a cube.

Since $\gamma/\alpha=-1/(64\alpha^2)$ and $-1/64=(-1/4)^3$, the ratio $\gamma/\alpha$ is a cube exactly when $\alpha$ is a cube. We may therefore count using
\begin{equation}
\label{eq:r-param}
  r=\frac{b+s}{b-s}=\frac{\gamma}{\alpha}.
\end{equation}
Replacing $s$ by $-s$ inverts $r$ and leaves the cube condition unchanged.

Conversely, solving \eqref{eq:r-param} gives
  $s=b\frac{r-1}{r+1}$.
Substituting $s^2=b^2+1/4$ yields
\begin{equation}
\label{eq:b2-from-r}
  b^2=-\frac{(r+1)^2}{16r}.
\end{equation}
Hence an admissible $b\in\F_q^*$ exists precisely when $r\ne\pm1$ and $-1/r$ is a square in $\F_q^*$.  The value $r=1$ gives $s=0$, i.e. the zero-discriminant case, while $r=-1$ gives $b=0$ by \eqref{eq:b2-from-r}. Both are excluded here.

For each admissible $r$, Equation~\eqref{eq:b2-from-r} gives the two values $\pm b$.  For each fixed $b$, the two choices $s$ and $-s$ give $r$ and $r^{-1}$.  Since $r=\pm1$ has been excluded, these are distinct.  Therefore the number of admissible values of $b$ is exactly the number of admissible values of $r$.

If $q\equiv1\pmod{12}$, then $-1$ is a square, and the condition $-1/r\in(\F_q^*)^2$ is equivalent to $r\in(\F_q^*)^2$.  Together with the cube condition, this says that $r$ is a sixth power.  There are $(q-1)/6$ sixth powers in $\F_q^*$, and both $r=1$ and $r=-1$ are sixth powers in this congruence class.  Hence
  $C_q=\frac{q-1}{6}-2=\frac{q-13}{6}$.

If $q\equiv7\pmod{12}$, then $-1$ is a nonsquare, and the condition $-1/r\in(\F_q^*)^2$ is equivalent to $r$ being a nonsquare.  Thus $r$ must be a nonsquare cube.  There are $(q-1)/6$ nonsquare cubes.  In this congruence class $r=-1$ is a nonsquare cube and must be excluded, while $r=1$ is a square and is not counted.  Hence
 $ C_q=\frac{q-1}{6}-1=\frac{q-7}{6}$.

Finally, in the nonzero-square-discriminant case, Theorem~\ref{ST} permits only factorization types $(1,1,1)$ and $(3)$ for a cubic over $\F_q$.  Thus every nonzero-square-discriminant value of $b$ not counted above gives no root of $G_b$ in $\F_q$.
\end{proof}

We now determine the root distribution of the cubic $G_b$ as $b$ varies in $\F_q^*$.
\begin{proposition}
\label{prop:subcase-1a-filled}
Assume $q\equiv1\pmod 6$.  For $b\in\F_q^*$, the number of distinct roots of $G_b$ in $\F_q$ has the following distribution.
\[
\begin{array}{c|cccc}
 &0&1&2&3\\ \hline
q\equiv1\pmod{12} &(q-1)/3 &(q-1)/2 &2 &(q-13)/6\\[2mm]
q\equiv7\pmod{12} &(q-1)/3 &(q+1)/2 &0 &(q-7)/6.
\end{array}
\]
\end{proposition}

\begin{proof}
The discriminant of $G_b$ is $-\frac{27}{16}\delta_b$, and $-3$ is a square in $\F_q$ because $q\equiv1\pmod3$.  Therefore $G_b$ has square, nonsquare, or zero discriminant exactly according as $\delta_b$ is a nonzero square, a nonsquare, or zero.

If $\delta_b$ is a nonsquare, then the discriminant of $G_b$ is a nonsquare.  Stickelberger's parity theorem gives factorization type $(1,2)$, so $G_b$ has exactly one root in $\F_q$.  By Lemma~\ref{lem:N1N2}, this occurs for $(q-1)/2$ values of $b$ when $q\equiv1\pmod{12}$ and for $(q+1)/2$ values of $b$ when $q\equiv7\pmod{12}$.

If $\delta_b$ is a nonzero square, then Theorem~\ref{thm:cubic-count} shows that $G_b$ splits completely for $(q-13)/6$ values of $b$ when $q\equiv1\pmod{12}$ and for $(q-7)/6$ values when $q\equiv7\pmod{12}$.  The remaining nonzero-square-discriminant cases have no root in $\F_q$.  Since the numbers of nonzero-square values of $\delta_b$ are $(q-5)/2$ and $(q-3)/2$, respectively, the corresponding zero-root counts are
$\frac{q-5}{2}-\frac{q-13}{6}=\frac{q-1}{3},  \frac{q-3}{2}-\frac{q-7}{6}=\frac{q-1}{3}.$

Finally, $\delta_b=0$ is possible exactly when $-1$ is a square in $\F_q$, which in the present setting is $q\equiv1\pmod{12}$.  Then there are exactly two such values of $b$.  For them $b^2=-1/4$, and
$G_b(X)=(X+b)^2(X-2b),$
so $G_b$ has exactly two distinct roots, namely $-b$ and $2b$.  This proves the table.
\end{proof}

\begin{proposition}
\label{prop:subcase-1b-filled}
Assume $q\equiv1\pmod6$, $b\in\F_q^*$, and $\delta_b \ne0$.  The branch
$X\notin\F_q, \  X^{q+1}=-\frac14$
contributes exactly two solutions if $\delta_b$ is a nonsquare in $\F_q$, and contributes no solution if $\delta_b$ is a square in $\F_q$.
\end{proposition}

\begin{proof}
On the branch $X^{q+1}=-1/4$, Equation~\eqref{eq:main-equation} becomes
 $ X^q+X+2b=0$.
  Multiplying by $X$ and using $X^{q+1}=-1/4$ gives
\begin{equation}
\label{eq:quadratic-filled}
  X^2+2bX-\frac14=0.
\end{equation}
Its discriminant is $\delta_b$.  If $\delta_b$ is a square in $\F_q$, then both roots of \eqref{eq:quadratic-filled} lie in $\F_q$, so they do not belong to the present branch.  If $\delta_b$ is a nonsquare in $\F_q$, then \eqref{eq:quadratic-filled} has two roots in $\F_{q^2}\setminus\F_q$.  Since its coefficients lie in $\F_q$, the two roots are $X$ and $X^q$.  Their sum is $-2b$ and their product is $-1/4$, so each root satisfies both $X^q+X+2b=0$ and $X^{q+1}=-1/4$.  Hence the two roots are precisely the two desired solutions.
\end{proof}

\begin{proposition}
\label{prop:subcase-1c-filled}
Assume $q\equiv1\pmod6$ and $b\in\F_q^*$.  The overlap branch
$X\in\F_q, \  X^{q+1}=X^2=-\frac14$
occurs if and only if $q\equiv1\pmod{12}$ and $\delta_b=0$.  For the two such values of $b$, Equation~\eqref{eq:main-equation} has exactly two distinct solutions in $\F_{q^2}$.
\end{proposition}

\begin{proof}
The equation $X^2=-1/4$ has roots in $\F_q$ if and only if $-1$ is a square in $\F_q$, which in the present congruence class is equivalent to $q\equiv1\pmod{12}$.  If $X\in\F_q$ and $X^2=-1/4$, then \eqref{eq:main-equation} becomes $2X+2b=0$, so $X=-b$ and therefore $\delta_b=1+4b^2=0$.  Conversely, if $\delta_b=0$, then $b^2=-1/4$, and
$G_b(X)=(X+b)^2(X-2b).$
Thus the subfield branch has the two distinct roots $-b$ and $2b$.  Among them, only $-b$ lies on the norm branch $X^{q+1}=X^2=-1/4$. Indeed, $(2b)^2=4b^2=-1\ne -1/4$.  The norm-branch quadratic \eqref{eq:quadratic-filled} has the single root $-b$ in this case, so it gives no further nonsubfield root.  Therefore \eqref{eq:main-equation} has exactly two distinct solutions.
\end{proof}

The three preceding propositions determine the subfield contribution. The nonsubfield norm branch contributes two roots exactly in the nonsquare-discriminant case, and the overlap calculation accounts for the zero-discriminant case.

\begin{theorem}\label{thm:subfield-theorem}
Let $q\equiv1\pmod6$.  For $b\in\F_q^*$, the number of solutions in $\F_{q^2}$ of
$4X^{q+2}+X^q+2X+2b=0$
has the following distribution.
\[
\begin{array}{c|cccccc}
 & n_0^{(q)}&n_1^{(q)}&n_2^{(q)}&n_3^{(q)}&n_4^{(q)}&n_5^{(q)}\\ \hline
q\equiv1\pmod{12} &(q-1)/3&0&2&(2q-8)/3&0&0\\[1mm]
q\equiv7\pmod{12} &(q-1)/3&0&0&2(q-1)/3&0&0.
\end{array}
\]
Here $n_i^{(q)}$ denotes the number of elements $b\in\F_q^*$ for which the displayed equation has exactly $i$ solutions.
\end{theorem}
\begin{proof}
If $\delta_b$ is a nonsquare, Propositions~\ref{prop:subcase-1a-filled} and~\ref{prop:subcase-1b-filled} give one subfield root and two nonsubfield roots, respectively. If $\delta_b$ is a nonzero square, there are three roots for $C_q$ parameters and none for the other $N_1-C_q$ parameters. If $\delta_b=0$, Proposition~\ref{prop:subcase-1c-filled} gives two roots for exactly two parameters, possible only when $q\equiv1\pmod{12}$. Thus the zero-root count is $N_1-C_q=(q-1)/3$, the three-root count is $N_2+C_q$, and the two-root count is two or zero according to the congruence class. Substituting the values of $N_2$ and $C_q$ gives the asserted distribution. These cases exhaust~$\F_q^*$.
\end{proof}

\section{The nonsubfield parameters}
\label{sec:nonsubfield}

Assume now that $b\in\F_{q^2}\setminus\F_q$.  Let the minimal polynomial of $b$ over $\F_q$ be
\begin{equation}
\label{eq:minpoly-b}
  T^2+AT+B,
  \qquad A,B\in\F_q.
\end{equation}
Thus
$b^q+b=-A$ and $B=b^{q+1}.$
The condition $b\notin\F_q$ is equivalent to
 $ D=A^2-4B\in C_1$, where $C_1$ is the set of nonsquares in $\F_q^*$ and $C_0$ denotes the set of nonzero squares. Write $X=u+vb$ with $u,v\in\F_q$.  Substitution into \eqref{eq:main-equation}, followed by separation of the $1$- and $b$-coordinates in the basis $\{1,b\}$, gives
\begin{equation}
\label{eq:uv-system}
\begin{cases}
  4u^3+4Buv^2-4Au^2v+3u-Av=0,\\
  -4Auv^2+4u^2v+4Bv^3+v+2=0.
\end{cases}
\end{equation}
Let
$\mathcal D'=\{(A,B)\in\F_q^2:A^2-4B\in C_1\},$
then for $(A,B)\in\mathcal D'$, we define
\begin{equation*}
  N(A,B)=\#\{(u,v)\in\F_q^2:\text{\eqref{eq:uv-system} holds}\}.
\end{equation*}

Observe that, by multiplying the first equation in System~\eqref{eq:uv-system} by $v$, the second by $u$, and then subtracting the resulting equations, we obtain
\begin{equation}
\label{eq:key-linear}
  -Av^2+2u(v-1)=0.
\end{equation}
This identity separates the analysis into the branch $v\ne1$ and the exceptional branch $v=1$. We write $m_i=\#\{(A,B)\in\mathcal D':N(A,B)=i\}$ for the coefficient-pair frequencies.

\begin{proposition}
\label{prop:quintic-reduction}
Let $b\in\F_{q^2}\setminus\F_q$ have minimal polynomial \eqref{eq:minpoly-b}.  Let $X=u+vb$ with $u,v\in\F_q$.  Every solution of \eqref{eq:main-equation} with $v\ne1$ satisfies
\begin{equation}
\label{eq:u-from-v}
  u=\frac{Av^2}{2(v-1)}
\end{equation}
and its $v$-coordinate is a root of
\begin{equation}
\label{eq:Q-AB}
  AQ_{A,B}(V)=(A^3-4AB)V^5-2(A^3-4AB)V^4-(4AB+A)V^3+3AV-2A=0.
\end{equation}
Equivalently, after division by $A$ when $A \ne0$, the same equation is
\begin{equation}
\label{eq:Q-main}
 Q_{A,B}(V)= (A^2-4B)V^5-2(A^2-4B)V^4-(4B+1)V^3+3V-2=0.
\end{equation}
In terms of $b$ and $b^q$, this is
\begin{equation}
\label{eq:Q-b}
  (b-b^q)^2V^5-2(b-b^q)^2V^4-(4b^{q+1}+1)V^3+3V-2=0.
\end{equation}
\end{proposition}

\begin{proof}
The identity \eqref{eq:key-linear} gives \eqref{eq:u-from-v} when $v\ne1$.  Substituting \eqref{eq:u-from-v} into the first equation of \eqref{eq:uv-system} and simplifying gives $-AvQ_{A,B}(v)/(2(v-1)^3)=0$. The value $v=0$ is impossible in the second equation of \eqref{eq:uv-system}, whose constant term is $2$. Thus \eqref{eq:Q-AB} follows. For $A\ne0$, substitution into the second equation gives $-Q_{A,B}(v)/(v-1)^2=0$. Conversely, every root of $Q_{A,B}$ with $v\ne0,1$, together with \eqref{eq:u-from-v}, satisfies both equations. This proves the required equivalence, as well as \eqref{eq:Q-main}.  Finally,
$(b-b^q)^2=(b+b^q)^2-4b^{q+1}=A^2-4B$ and  $b^{q+1}=B,$
so \eqref{eq:Q-main} is exactly \eqref{eq:Q-b}.
\end{proof}

\begin{proposition}
\label{prop:v-one}
If $v=1$ in \eqref{eq:uv-system}, then $A=0$.  Conversely, assume that
$A=0$ and $v=1$.  Then the solutions in this branch satisfy
$4u^2+4B+3=0.$
Thus, for an admissible pair $(0,B)\in\mathcal D'$, it contributes exactly
$1+\eta(-4B-3)$
solutions $u\in\F_q$.  Equivalently, the contribution is two if
$-4B-3\in C_0$, zero if $-4B-3\in C_1$, and one if $B=-3/4$.  The special
case $B=-3/4$ is admissible precisely when $3\in C_1$. Under
$q\equiv1\pmod 6$, this is equivalent to $q\equiv7\pmod{12}$.
\end{proposition}

\begin{proof}
Substituting $v=1$ into \eqref{eq:key-linear} gives
$  -A\cdot 1+2u(1^2-1)=0$, and hence $A=0$.

Conversely, assume $A=0$ and put $v=1$ in System~\eqref{eq:uv-system}.  The two
equations become
$4u^3+4Bu+3u=0$ and $4u^2+4B+3=0$, respectively.
The first equation factors as $ u(4u^2+4B+3)=0$.
Therefore every solution of the second equation also satisfies the first, and
the branch is equivalent to  $4u^2+4B+3=0$,
or equivalently $ u^2=-B-\frac34$.
Since $4$ is a square in $\F_q$, the number of solutions is
$ 1+\eta(-4B-3)$,
with the convention $\eta(0)=0$.  Thus there are two solutions when $-4B-3\in C_0$, no solutions when $-4B-3\in C_1$, and the unique solution $u=0$ when $-4B-3=0$, i.e.\ when $B=-3/4$.

It remains only to record the admissibility condition for the special
one-solution case.  Since $A=0$, the polynomial
$ T^2+AT+B=T^2+B$ is irreducible over $\F_q$ exactly when its discriminant $ D=A^2-4B=-4B$ lies in $C_1$.  For $B=-3/4$, this condition becomes $3\in C_1$.  Because
$q\equiv1\pmod 6$, the quadratic character of $3$ is determined by
$q\pmod{12}$. Namely, $3\in C_0$ when $q\equiv1\pmod{12}$ and
$3\in C_1$ when $q\equiv7\pmod{12}$.  Hence the $v=1$, $B=-3/4$ branch
contributes one solution when $q\equiv7\pmod{12}$ and is inadmissible when
$q\equiv1\pmod{12}$.
\end{proof}

\begin{proposition}
\label{prop:A-zero}
Assume $A=0$.  Then $B \ne0$, and the solutions of \eqref{eq:uv-system} are the disjoint union of the branch
$v=1,   4u^2+4B+3=0,$
and the branch
$u=0,  v\in\F_q\setminus\{1\},  4Bv^3+v+2=0.$
Consequently
\begin{equation}
\label{eq:A-zero-count}
  N(0,B)=\#\{v\in\F_q\setminus\{1\}:4Bv^3+v+2=0\}+1+\eta(-4B-3),
\end{equation}
and in particular $N(0,B)\le5$.
\end{proposition}

\begin{proof}
Since $A=0$ and $A^2-4B\in C_1$, one has $B \ne0$.  Equation~\eqref{eq:key-linear} becomes $u(v-1)=0$.  If $v=1$, Proposition~\ref{prop:v-one} gives the first branch.  If $v\ne1$, then $u=0$.  Substitution into \eqref{eq:uv-system} leaves the single equation
$4Bv^3+v+2=0.$
The two branches are disjoint, the cubic has at most three roots, and the quadratic branch has at most two roots.  This proves \eqref{eq:A-zero-count} and the bound.
\end{proof}

The exact nonsubfield contribution to the boomerang spectrum requires the fiber distribution
\begin{equation*}
N(A,B)=\begin{cases}
\#\{v\in\F_q\setminus\{0,1\}:Q_{A,B}(v)=0\},& A\ne0,\\[1mm]
\#\{v\in\F_q\setminus\{1\}:4Bv^3+v+2=0\}+1+\eta(-4B-3),& A=0.
\end{cases}
\end{equation*}
In the first line $Q_{A,B}$ denotes the normalized quintic in \eqref{eq:Q-main}. In the second line the formula is exactly \eqref{eq:A-zero-count}.  Together with the constraint $A^2-4B\in C_1$, this gives the exact finite-field distribution attached to the nonsubfield branch.

Next, we show that $N(A,B)=4$ is never attained for $(A,B)\in\mathcal D'$ when $q\equiv 1\pmod{12}$, and, when $q\equiv 7\pmod{12}$, we characterise the pairs attaining it by means of a residual cubic.

\begin{theorem}
\label{thm:N4-zero}
Let $q\equiv1\pmod{12}$.  Then $N(A,B)\ne4$ for all $(A,B)\in\mathcal D'$.  Consequently $m_4=0$ and the nonsubfield spectrum values lie in $\{0,1,2,3,5\}$.
\end{theorem}

\begin{proof}
We treat the cases $A=0$ and $A\ne0$ separately.

Assume first that $A=0$.  By Proposition~\ref{prop:A-zero},
$N(0,B)=\rho_B+1+\rho'_B,$
where
$\rho_B=\#\{v\in\F_q\setminus\{1\}:4Bv^3+v+2=0\},   \rho'_B=\eta(-4B-3).$
Since $(A,B)\in\mathcal D'$, here $D=A^2-4B=-4B\in C_1$. In particular $B\ne0$.  The only way a cubic over $\F_q$ can have exactly two distinct $\F_q$-roots is for it to have a repeated root, because if two distinct linear factors divide the cubic, the remaining linear factor is also defined over $\F_q$, and either it gives a third distinct root or it coincides with one of the two roots.

The discriminant of $4Bv^3+v+2$ is $-16B(108B+1)$.  Because $B\ne0$, the only admissible repeated-root candidate is
$B=-\frac1{108}.$
For this value,
$D=-4B=\frac1{27}=3^{-3}.$
Since $q\equiv1\pmod {12}$, the element $3$ is a square in $\F_q$, so $D$ is a square, contradicting $D\in C_1$.  Therefore the cubic is squarefree for every admissible $B$.  A squarefree cubic has $0$, $1$, or $3$ roots in $\F_q$, never exactly $2$.

It remains to check the exceptional quadratic branch.  If $-4B-3=0$, then $D=-4B=3$, again a square when $q\equiv1\pmod {12}$, contradicting $D\in C_1$.  Hence $-4B-3\ne0$, and the branch $v=1$ contributes either $0$ or $2$ roots.  Combining a contribution in $\{0,1,3\}$ from the cubic branch with a contribution in $\{0,2\}$ from the exceptional branch gives only
$0,1,2,3,5,$
so $N(0,B)\ne4$.

Assume now that $A\ne0$.  Then the exceptional branch $v=1$ does not occur, because Proposition~\ref{prop:v-one} shows that $v=1$ forces $A=0$.  Thus $N(A,B)$ is the number of distinct roots in $\F_q$ of the quintic $Q_{A,B}$ in \textup{\ref{eq:Q-main}}.  A degree-$5$ polynomial over $\F_q$ with exactly four distinct $\F_q$-roots must have a repeated $\F_q$-root, because after division by the four distinct linear factors over $\F_q$, the remaining quotient is linear over $\F_q$, and its root must coincide with one of the four roots.

Let $v_0$ be such a repeated root.  The values $v_0=0$ and $v_0=1$ are impossible, since $Q_{A,B}(0)=-2$ and $Q_{A,B}(1)=-A^2$.  Write $S=A^2$ and $D=A^2-4B$.  Then
$Q_{A,B}(V)=DV^5-2DV^4-(S-D+1)V^3+3V-2.$
The equations $Q_{A,B}(v_0)=0$ and $Q'_{A,B}(v_0)=0$ are
\begin{align*}
  Dv_0^5-2Dv_0^4-(S-D+1)v_0^3+3v_0-2&=0,\\
  5Dv_0^4-8Dv_0^3-3(S-D+1)v_0^2+3&=0.
\end{align*}
Multiplying the derivative equation by $v_0/3$ and subtracting the first equation eliminates $S-D+1$ and gives
$\frac23D v_0^5-\frac23D v_0^4-2v_0+2=0.$
Equivalently,
$\frac{2}{3}(v_0-1)(Dv_0^4-3)=0.$
Since $v_0\ne1$, we obtain $D=3v_0^{-4}$.  Therefore
$\eta(D)=\eta(3)\eta(v_0^{-4})=\eta(3)=1,$
because $q\equiv1\pmod {12}$ implies that $3$ is a square in $\F_q$.  This contradicts the defining condition $D=A^2-4B\in C_1$ for $(A,B)\in\mathcal D'$.  Hence the quintic cannot have exactly four distinct $\F_q$-roots.
\end{proof}

\begin{theorem}
\label{thm:N4-zero2}
Let $q\equiv7\pmod{12}$, and let $(A,B)\in\mathcal D'$.  The following hold.
\begin{enumerate}
\item[$(1)$] If $A=0$, then
$N(0,B)=4\quad\Longleftrightarrow\quad B=-\frac{1}{108}\quad\text{and}\quad \eta(5)=1.$
\item[$(2)$] If $A\ne0$, then $N(A,B)=4$ if and only if there exists $v\in\F_q\setminus\{0,1\}$ such that
\begin{equation}
\label{eq:N4-q7-param}
  A^2=-\frac{(v-1)^3(v+3)}{v^4},
  \qquad
  B=-\frac{v^3-6v+8}{4v^3},
\end{equation}
and the cubic
\begin{equation*}
  R_v(V)=3V^3+6(v-1)V^2+(3v^2-4v)V-2v^2
\end{equation*}
has three distinct roots in $\F_q$.
\end{enumerate}
Consequently,
\begin{equation*}
  m_4=\frac{1+\eta(5)}{2}+2|\mathcal V_q|,
\end{equation*}
where
\begin{equation*}
\mathcal V_q=\bigl\{v\in\F_q\setminus\{0,1,-3\}:\eta((v-1)(v+3))=-1\text{ and }R_v\text{ has three distinct roots in }\F_q\bigr\}.
\end{equation*}
\end{theorem}

\begin{proof}
Since $q\equiv7\pmod{12}$, one has $\eta(-1)=-1$ and $\eta(3)=-1$.

Assume first that $A=0$.  Proposition~\ref{prop:A-zero} gives
$N(0,B)=\#\{v\in\F_q\setminus\{1\}:4Bv^3+v+2=0\}+1+\eta(-4B-3).$

If $-4B-3=0$, then $B=-3/4$.  The exceptional branch $v=1$ contributes the unique solution $u=0$.  Moreover
$4Bv^3+v+2=-3v^3+v+2=-(v-1)(3v^2+3v+2).$
Since the cubic branch excludes $v=1$, it contributes at most the two roots of $3v^2+3v+2$.  Hence this case gives $N(0,B)\le3$, and so it cannot give $N(0,B)=4$.

Now assume $-4B-3\ne0$.  Then $1+\eta(-4B-3)$ is either $0$ or $2$.  The cubic branch contributes at most three roots, and hence, if $N(0,B)=4$, the exceptional branch must contribute $2$ and the cubic branch must contribute exactly two distinct roots.  As above, a cubic over $\F_q$ has exactly two distinct $\F_q$-roots only if it has a repeated root.  The discriminant of $4Bv^3+v+2$ is $-16B(108B+1)$.  Since $D=-4B\in C_1$, we have $B\ne0$, and the only possible repeated-root value is
$B=-\frac1{108}.$
For this value,
$D=-4B=\frac1{27}=3^{-3},$
which is a nonsquare because $\eta(3)=-1$.  Also
$4Bv^3+v+2=-\frac1{27}(v+3)^2(v-6),$
so the cubic branch contributes exactly the two distinct roots $-3$ and $6$.  Finally,
$\eta(-4B-3)=\eta\left(-\frac{80}{27}\right) =\eta(-1)\eta(5)\eta(3)^{-3}=\eta(5),$
using $80=2^4\cdot5$ and the fact that powers of $2$ and fourth powers are squares.  Thus
$N(0,-1/108)=2+1+\eta(5),$
which equals $4$ exactly when $\eta(5)=1$.  This proves part \textup{(1)}.

Assume now that $A\ne0$.  Then the exceptional branch $v=1$ does not occur, and $N(A,B)$ is the number of distinct $\F_q$-roots of the normalized quintic $Q_{A,B}$ in \eqref{eq:Q-main}.  If $N(A,B)=4$, the same degree-$5$ argument used in the proof of Theorem~\ref{thm:N4-zero} shows that $Q_{A,B}$ has a repeated root $v\in\F_q$.  Since $Q_{A,B}(0)=-2$ and $Q_{A,B}(1)=-A^2$, this repeated root satisfies $v\in\F_q\setminus\{0,1\}$.

Put $S=A^2$ and $D=A^2-4B$.  From $Q_{A,B}(v)=Q'_{A,B}(v)=0$, the same elimination gives
$-\frac{2}{3}(v-1)(Dv^4-3)=0.$
Thus $D=3/v^4$.  Substituting this into
$Q_{A,B}(v)=Dv^5-2Dv^4-(S-D+1)v^3+3v-2=0$
and solving for $S$ gives
$S=A^2=-\frac{(v-1)^3(v+3)}{v^4}.$
Since $B=(S-D)/4$, we also get
$B=-\frac{v^3-6v+8}{4v^3}.$
These are exactly the formulas in \eqref{eq:N4-q7-param}.  The condition $A\ne0$ excludes $S=0$, i.e. $v=-3$. The exclusions $v=0,1$ have already been obtained.

Conversely, suppose that $v\in\F_q\setminus\{0,1,-3\}$ and that $A^2$ and $B$ are given by \eqref{eq:N4-q7-param}.  Then $A\ne0$, and
$D=A^2-4B=\frac{3}{v^4}.$
Substitution into $Q_{A,B}(V)$ gives the factorization
\begin{equation*}
  Q_{A,B}(V)=\frac{(V-v)^2}{v^4}R_v(V),
\end{equation*}
where
$R_v(V)=3V^3+6(v-1)V^2+(3v^2-4v)V-2v^2.$
A direct evaluation gives
$R_v(v)=12v^2(v-1)\ne0,$
so the double root $v$ is not a root of the residual cubic.  The discriminant of the residual cubic is
$\disc(R_v)=-48v^2(v+3)(9v^2-12v+8).$
For $q\equiv7\pmod{12}$, the quadratic factor $9v^2-12v+8$ has discriminant $-144$, which is a nonsquare because $\eta(-1)=-1$.  Hence this factor has no root in $\F_q$.  Thus, for $v\ne0,-3$, the residual cubic is squarefree.  Moreover $Q_{A,B}(0)=-2$ and $Q_{A,B}(1)=-A^2\ne0$, so no root counted by $R_v$ can be $0$ or $1$ in a way that would be excluded from the quintic count.  Therefore $Q_{A,B}$ has exactly four distinct roots in $\F_q$ if and only if $R_v$ has three distinct roots in $\F_q$.

Here $D=3/v^4$ and $\eta(3)=-1$, so $D\in C_1$.
The additional condition that $A^2$ be a nonzero square is
$\eta\left(-\frac{(v-1)^3(v+3)}{v^4}\right)=1.$
Since $v^4$ is a square and $\eta(-1)=-1$, this is equivalent to
$\eta((v-1)(v+3))=-1.$
Thus the admissible repeated-root parameters are precisely the elements of $\mathcal V_q$ in the theorem.  Finally, if $N(A,B)=4$, the repeated root is unique because a degree-$5$ polynomial with four distinct roots has one double root and three simple roots.  Hence the map from the $N=4$ parameters with $A\ne0$ to their repeated-root parameter $v$ is one-to-one after passing to $A^2$.  Each admissible value of $A^2$ gives exactly two choices $A=\pm\sqrt{A^2}$, and hence two coefficient pairs $(A,B)$.  This gives the contribution $2|\mathcal V_q|$, while part \textup{(1)} gives the trace-zero contribution $(1+\eta(5))/2$.
\end{proof}

\section{A character-sum bound for the nonsubfield quintic}
\label{sec:incidence-bound}
Here and below, the phrase \emph{nonzero-trace branch} means the branch in which $b+b^q\neq0$.  Since $A=-(b+b^q)$, this is exactly the case $A\neq0$. The complementary trace-zero branch is $A=0$.

Recall Proposition \ref{prop:quintic-reduction}. The coefficients of \eqref{eq:Q-main} lie in $\F_q$.  This permits the following normalization in the nonzero-trace branch ($A\neq0$).  Write
$D=A^2-4B$ as before.
For $A \ne0$, the values $v=0$ or $v=1$ cannot occur in \eqref{eq:Q-main}, because $Q_{A,B}(0)=-2$ and $Q_{A,B}(1)=-A^2$ in the normalized form \eqref{eq:Q-main}.  Since $4B=A^2-D$, Equation~\eqref{eq:Q-main} becomes
$Dv^5-2Dv^4-(A^2-D+1)v^3+3v-2=0.$
Equivalently,
\begin{equation}
\label{eq:D-as-rational}
  D=\frac{(A^2+1)v^3-3v+2}{v^3(v-1)^2}.
\end{equation}
Thus for $A \neq 0$, condition $D\in C_1$ becomes a quadratic-character condition on a polynomial in $v$ for $v$ satisfying $Q_{A,B}(v)=0$.

\begin{theorem}
\label{thm:incidence-bound}
Assume $A \ne0$ and put
  $P_A(V)=V\bigl((A^2+1)V^3-3V+2\bigr)\in\F_q[V]$.
For fixed $A\in\F_q^*$, let $I_A$ be the number of pairs $(B,v)$ such that $A^2-4B\in C_1$, $v\in\F_q\setminus\{0,1\}$, and $Q_{A,B}(v)=0$.
Let $Z_A=\#\{v\in\F_q\setminus\{0,1\}:(A^2+1)v^3-3v+2=0\}$. Then
\begin{equation}
\label{eq:IA-exact-main}
  I_A=\frac{q-1-Z_A}{2}-\frac12\sum_{v\in\F_q}\eta(P_A(v)).
\end{equation}
Consequently,
\begin{equation}
\label{eq:IA-bound-main}
  \left|I_A-\frac{q-1-Z_A}{2}\right|\le \frac32\sqrt q,
\end{equation}
and hence
$\frac{q-4}{2}-\frac32\sqrt q \le I_A\le \frac{q-1}{2}+\frac32\sqrt q .$
\end{theorem}

\begin{proof}
Fix $A\in\F_q^*$ throughout the proof.  For each fixed
$v\in\F_q\setminus\{0,1\}$, the denominator in
\eqref{eq:D-as-rational} is nonzero, and therefore the displayed rational expression determines exactly one value of $D$.  This in turn determines exactly one value of
$B=(A^2-D)/4$.  Thus the correspondence used here is a correspondence from a value of $v$ to the incidence $(B,v)$.  We do not require the map $v\mapsto D$ to be injective. If two different values of $v$ produce the same value of $D$, they merely give two incidences with the same parameter $B$, and $I_A$ is defined to count incidences $(B,v)$ rather than distinct parameters $B$.
This value corresponds to a nonsubfield parameter exactly when $D\in C_1$.  Since $(v-1)^2$ is a square and $\eta(v^3)=\eta(v)$, the condition $D\in C_1$ is equivalent to
$\eta(P_A(v))= \eta\left(v\bigl((A^2+1)v^3-3v+2\bigr)\right)=-1.$
This is precisely $\eta(P_A(v))=-1$.

The character filter gives
\begin{align*}
I_A&= \#\bigl\{(B,v) \in \F_q \times (\F_q \setminus \{0,1\}) \,:\, Q_{A,B}(v) = 0 \text{ and } A^2 - 4B \in C_1\bigr\}\\
& =\sum_{v\in\F_q\setminus\{0,1\}}\frac{1-\eta(P_A(v))}{2}-\frac{Z_A}{2}.
\end{align*}

The subtraction by $Z_A/2$ removes the values for which the numerator in \eqref{eq:D-as-rational} is zero. For such values the formula gives $D=0$, which is not a nonsquare.  Since $P_A(0)=0$ and $P_A(1)=A^2$, we have
$\sum_{v\in\F_q\setminus\{0,1\}}\eta(P_A(v)) =\sum_{v\in\F_q}\eta(P_A(v))-1.$
Therefore
\[
  I_A=\frac{q-2}{2}-\frac12\left(\sum_{v\in\F_q}\eta(P_A(v))-1\right)-\frac{Z_A}{2}
  =\frac{q-1-Z_A}{2}-\frac12\sum_{v\in\F_q}\eta(P_A(v)).
\]
This proves \eqref{eq:IA-exact-main}.

It remains to bound the character sum.  The polynomial $P_A(V)$ is not of the form $cH(V)^2$, because $V=0$ is a simple root.  Its degree is at most four.  By \Cref{thm:weil-polynomial-character}, applied to $P_A$,
$\left|\sum_{v\in\F_q}\eta(P_A(v))\right|\le3\sqrt q.$
Also $Z_A\le3$, because $(A^2+1)V^3-3V+2$ has degree at most three.  The estimates \eqref{eq:IA-bound-main} and the displayed two-sided bound follow.
\end{proof}

The bound above is an incidence bound.  It controls the total number of admissible roots of the quintic $Q_{A,B}$ with $A\neq0$ as $B$ varies with fixed $A$.  

\section{Rational curves and fiber moments}
\label{sec:geometry}

The rational expression~\eqref{eq:D-as-rational} associates the fixed-trace character sums with a plane curve. The first curve 
describes the vanishing of the numerator in \eqref{eq:D-as-rational}, which forces $D=0$ and lies outside the admissible nonsubfield domain.  We first describe an explicit birational parametrization and the singularities of its projective closure and then use a second rational curve to obtain
exact secant and residual-cubic formulas for the higher moments.

Recall from~\eqref{eq:D-as-rational} that every $\F_q$-root $v$ of the quintic $Q_{A,B}$, with $v \ne0,1$, satisfies
\begin{equation}\label{Qroot}
  D = \frac{(A^2+1)v^3-3v+2}{v^3(v-1)^2},
  \qquad D=A^2-4B.
\end{equation}
Clearing denominators and using $v^3-3v+2=(v-1)^2(v+2)$, Equation~\eqref{eq:D-as-rational} is equivalent to
\begin{equation*}
  \frac{A^2}{(v-1)^2}=D-\frac{v+2}{v^3}.
\end{equation*}
For $A\ne0$, the right-hand side is a nonzero square. This character restriction will reappear in the secant count.

Write $\overline{\F}_q$ for an algebraic closure of $\F_q$, $\mathbb A^2$ for the affine plane, and $\mathbb P^2$ for the projective plane, whose points are nonzero triples modulo common nonzero scaling. A birational map is a rational map with a rational inverse on nonempty open subsets. A curve is rational over $\F_q$ when it is birational over $\F_q$ to the projective line.

The numerator-zero curve is the affine plane curve
\begin{equation*}
  \mathcal C:\quad (A^2+1)v^3-3v+2=0
  \quad\subset\mathbb A^2_{A,v}.
\end{equation*}
For $v\ne0,1$, this is the numerator-zero locus of \eqref{eq:D-as-rational}. Solving for $A^2$ gives
\begin{equation*}
  A^2=-\frac{(v-1)^2(v+2)}{v^3},
\end{equation*}
so that a point $(A,v)\in\mathcal C(\F_q)$ with $v\ne0,1$ exists if and only if $\eta(-v(v+2))\ne-1$.

\begin{proposition}
\label{prop:C-genus-zero}
The affine curve
\[
  \mathcal C:\quad (A^2+1)v^3-3v+2=0
\]
is rational over $\F_q$.  More precisely, the map
\begin{equation}
\label{eq:C-param-t}
  t\longmapsto \bigl(A(t),v(t)\bigr)
  =\left(\frac{t(t^2+3)}{2},-\frac{2}{t^2+1}\right)
\end{equation}
gives a parametrization of $\mathcal C$ on the open set $t^2+1\ne0$, and away from $v=1$ its inverse is
\begin{equation}
\label{eq:C-param-inverse}
  t=\frac{Av}{v-1}.
\end{equation}
The exceptional value $v=1$ is the point $(A,v)=(0,1)$, and it is obtained from $t^2=-3$.
\end{proposition}

\begin{proof}
Solving the equation of $\mathcal C$ for $A^2$ gives
\begin{equation}
\label{eq:A2-rational-C}
  A^2=-\frac{(v-1)^2(v+2)}{v^3}.
\end{equation}
For $v\ne1$, put $t=Av/(v-1)$.  Then \eqref{eq:A2-rational-C} gives
$t^2=\frac{A^2v^2}{(v-1)^2}=-\frac{v+2}{v}.$
Hence $v(t^2+1)=-2$, so $v=-2/(t^2+1)$.  Substituting this into $A=t(v-1)/v$ gives
$A=t\frac{v-1}{v} =t\frac{-2/(t^2+1)-1}{-2/(t^2+1)} =\frac{t(t^2+3)}{2}.$
Thus every point of $\mathcal C$ with $v\ne1$ is obtained from \eqref{eq:C-param-t}, and \eqref{eq:C-param-inverse} recovers $t$ from it.

Conversely, for $t^2+1\ne0$, set $A=A(t)$ and $v=v(t)$ as in \eqref{eq:C-param-t}.  The identity $t^2=-(v+2)/v$ holds by construction, and therefore
$A^2=\frac{t^2(v-1)^2}{v^2} =-\frac{(v-1)^2(v+2)}{v^3}.$
Multiplying by $v^3$ gives $(A^2+1)v^3-3v+2=0$.  Finally, if $v=1$, the curve equation gives $A=0$, and $v(t)=1$ is equivalent to $t^2=-3$, for which $A(t)=0$.
\end{proof}

We use the following standard notions for plane curves~\cite[Chapters~7--8]{Fulton}. The normalization is the nonsingular projective curve with the same function field. The arithmetic genus $p_a$ of a degree-$d$ plane curve is $(d-1)(d-2)/2$. At a singular point $P$, $\delta(P)$ is the dimension over the algebraically closed ground field of the quotient of the normalized local ring by the original local ring. An ordinary node has two distinct tangent lines and $\delta=1$, and it is split when both branches are defined over the ground field. The genus formula is $p_a=g+\sum_P\delta(P)$, where $g$ is the genus of the normalization. The value semigroup of a single branch consists of the orders of its nonzero local functions in a normalization parameter.

\begin{proposition}
\label{prop:C-projective-singularities}
Let $\overline{\mathcal C}$ be the projective closure of $\mathcal C$, defined by
$F(A,V,Z):=(A^2+Z^2)V^3-3VZ^4+2Z^5=0$ in $\mathbb P^2_{A,V,Z}.$
Then $\overline{\mathcal C}$ is an irreducible plane quintic.  Its arithmetic genus is $6$.  Its singular points over $\overline{\F}_q$ are
$P_*=[1:0:0],  P_0=[0:1:0], P_1=[0:1:1].$
The corresponding $\delta$-invariants are $4,1,1$, respectively. The point $P_*$ has one branch, with value semigroup $\langle3,5\rangle$.  The point $P_1$ is a split ordinary node over $\F_q$, while $P_0$ is split over $\F_q$ exactly when $-1$ is a square in $\F_q$.
\end{proposition}

\begin{proof}
Since $F$ is homogeneous of degree $5$, the arithmetic genus of this plane quintic is $(5-1)(5-2)/2=6$.  We first prove irreducibility.  In the affine chart $Z=1$, the equation is quadratic in $A$ and gives
$A^2=-\frac{(v-1)^2(v+2)}{v^3}$ in $\overline{\F}_q(v).$
The rational function on the right has odd valuation at $v=0$ and at $v=-2$, and therefore is not a square in $\overline{\F}_q(v).$ The coefficients, viewed as polynomials in $v$, have greatest common divisor one. Gauss's lemma therefore gives irreducibility of the affine equation from irreducibility over $\overline{\F}_q(v)$. Homogenization has no factor $Z$, so the projective curve is also irreducible.

The singular points are obtained by solving the equation $F(A,V,Z)=0$ together with its first partial derivatives
$2AV^3, 3V^2(A^2+Z^2)-3Z^4,  2ZV^3-12VZ^3+10Z^4.$
In the affine chart $Z=1$, this gives the unique singular point $(A,v)=(0,1)$.  On the line at infinity $Z=0$, the equation $F(A,V,Z)=0$ reduces to $A^2V^3=0$, and the derivative equations give exactly two singular points $[1:0:0]$ and $[0:1:0]$.

At $P_1=(0,1)$, which is the point $(A,V)=(0,1)$ in the chart $Z=1$, put $v=1+u$.  The equation becomes
$A^2+3u^2+\text{terms of total degree at least }3=0$. Since the quadratic part factors as $(A-\sqrt{-3}u)(A+\sqrt{-3}u)$, the point $P_1$ is an ordinary node.  Since $q\equiv1\pmod3$, the element $-3$ is a square in $\F_q$, so the two tangent lines are defined over $\F_q$, and hence this node is split over $\F_q$.  Its $\delta$-invariant is $1$.

At $P_0=[0:1:0]$, use the chart $V=1$, with local coordinates $a=A/V$ and $z=Z/V$.  The equation is
$a^2+z^2-3z^4+2z^5=0.$
The tangent cone is $a^2+z^2$.  Therefore $P_0$ is an ordinary node over $\overline{\F}_q$, with $\delta(P_0)=1$.  Its two tangent lines are defined over $\F_q$ exactly when $-1$ is a square in $\F_q$.

At $P_*=[1:0:0]$, use the chart $A=1$, with local coordinates $y=V/A$ and $z=Z/A$.  The local equation is
$(1+z^2)y^3-3yz^4+2z^5=0.$
The parametrization in Proposition~\ref{prop:C-genus-zero}, written near $t=\infty$ with $s=1/t$, gives
$z=2s^3+O(s^5), y=-4s^5+O(s^7).$
The parametrization extends to the projective line by
\[
[T:U]\longmapsto
[T(T^2+3U^2)(T^2+U^2):-4U^5:2U^3(T^2+U^2)].
\]
The coordinates have no common zero, and the affine inverse already proved shows that this map is birational. Hence it is the normalization map. The only point mapping to $P_*$ is $[1:0]$, namely $t=\infty$, so $P_*$ has one branch. Since the normalization has genus zero, the genus formula and the two node contributions give $\delta(P_*)=6-1-1=4$. The displayed orders $3$ and $5$ show that its value semigroup contains $\langle3,5\rangle$. This semigroup has the four gaps $1,2,4,7$. For a single branch, the number of gaps equals $\delta$. Indeed, at every order the normalized local ring has one coefficient over the ground field. The original local ring supplies that coefficient exactly when the order lies in its value semigroup. The missing orders therefore count the dimension of the quotient defining $\delta$. Hence there can be no additional semigroup values in those gaps, and the semigroup is exactly $\langle3,5\rangle$.
\end{proof}

To compute the frequencies $m_i$, we use moments of $N$. For $r \geq 0$, define
\[
M_r=\sum_{(A,B)\in\mathcal D'}N(A,B)^r,\qquad
\mathfrak F_r=\sum_{(A,B)\in\mathcal D'}(N(A,B))_r,
\]
where $(n)_r=n(n-1)\cdots(n-r+1)$ is the falling factorial, with $(n)_0=1$ and $N(A,B)^0=1$. In particular, $M_0=\#\mathcal D'=(q^2-q)/2$. For each $r$, the superscripts $0$ and $\ne0$ indicate restrictions to $A=0$ and $A\ne0$, respectively.

\begin{theorem}
\label{thm:first-moment-exact}
Let $q\equiv1\pmod6$ and put $\varepsilon=\eta(-1)$, equal to $1$ or $-1$ according as $q\equiv1$ or $7\pmod{12}$. Then
\begin{equation*}
M_1=\frac{q^2-2q+\varepsilon}{2},\qquad
\mathfrak F_1^0=q-\frac{3-\varepsilon}{2},\qquad
\mathfrak F_1^{\ne0}=\sum_{A\in\F_q^*}I_A=\frac{(q-1)(q-3)}2.
\end{equation*}
Moreover, the character sums and zero counts of Theorem~\textup{\ref{thm:incidence-bound}} satisfy
\begin{equation*}
\sum_{A\in\F_q^*}\sum_{v\in\F_q}\eta(P_A(v))=q+2+\varepsilon,
\qquad \sum_{A\in\F_q^*}Z_A=q-4-\varepsilon.
\end{equation*}
\end{theorem}

\begin{proof}
For each $X\in\F_{q^2}$ there is exactly one $b\in\F_{q^2}$ satisfying \eqref{eq:main-equation}. For $b\in\F_{q^2}^*$, put $N_F(b)=\#\{X\in\F_{q^2}:4X^{q+2}+X^q+2X+2b=0\}$. Hence $\sum_b N_F(b)=q^2$, where the definition of $N_F$ is extended to $b=0$. For $b=0$, \eqref{eq:factor-subfield} shows that every root is either in $\F_q$ or has norm $-1/4$. The subfield equation is $X(4X^2+3)=0$ and has three distinct roots because $-3$ is a square. On the other branch the equation becomes $X^q+X=0$. Together with the norm condition this forces $X^2=1/4$, and neither $X=1/2$ nor $X=-1/2$ has trace zero. Thus $N_F(0)=3$.

By Theorem~\ref{thm:subfield-theorem}, $\sum_{b\in\F_q^*}N_F(b)=\sum_{i}i n_i^{(q)}=2q-3-\varepsilon$. Every nonsubfield coefficient pair represents two output parameters. Therefore $2M_1=q^2-3-(2q-3-\varepsilon)$, proving the first formula.

Next, we compute $\mathfrak F_1^0$ and $\mathfrak F_1^{\ne0}$. For $A=0$ and $v\ne1$, Proposition~\ref{prop:A-zero} gives $u=0$ and $D=(v+2)/v^3$. This is admissible exactly when $\eta(v(v+2))=-1$. The quadratic-character sum $\sum_v\eta(v(v+2))=-1$ shows that there are $(q-1)/2$ such $v$ before excluding $v=1$. At $v=1$, we get $\eta(v(v+2))=\eta(3)=\varepsilon$, so $v=1$ was counted exactly when $\varepsilon=-1$. Excluding it therefore removes $(1-\varepsilon)/2$ values. The other branch contributes
\[
\sum_{D\in C_1}\bigl(1+\eta(D-3)\bigr)
=\frac{q-1}{2}+\frac12\left(\sum_{D\ne0}\eta(D-3)-\sum_D\eta(D(D-3))\right)
=\frac{q-1}{2},
\]
because the two sums in parentheses both equal $-1$, using $\eta(-3)=1$. Adding the branches gives $\mathfrak F_1^0=q-(3-\varepsilon)/2$. Subtracting this from $M_1$ gives $\mathfrak F_1^{\ne0}$.

For $v\ne0$, write $c_v=(v-1)^2(v+2)/v^3$. Then $\eta(P_A(v))=\eta(A^2+c_v)$. The values $v=1,-2$ give $c_v=0$, so each contributes $q-1$ to the sum over $A\ne0$. For the other $q-3$ values, the quadratic-character sum gives $\sum_{A\ne0}\eta(A^2+c_v)=-1-\eta(c_v)$. Also $\sum_{v\notin\{0,1,-2\}}\eta(c_v)=-1-\varepsilon$, since away from $v=1$ its character is $\eta(v(v+2))$. Consequently the aggregate is $2(q-1)-(q-3)+1+\varepsilon=q+2+\varepsilon$. Finally, summing \eqref{eq:IA-exact-main} over $A\ne0$ and inserting this aggregate and the value of $\sum_A I_A$ gives $\sum_A Z_A=q-4-\varepsilon$.
\end{proof}

The higher moments count ordered tuples of distinct roots in the same quintic fiber. The line-intersection description below gives explicit formulas for these counts.

For the branch $A\ne0$, put $S=A^2$ and $D=A^2-4B$.  Equation~\eqref{Qroot} is equivalent to
\begin{equation}
\label{eq:D-line-form}
  D=Sx(v)+y(v),
  \qquad
  x(v)=\frac{1}{(v-1)^2},
  \qquad
  y(v)=\frac{v+2}{v^3},
\end{equation}
for $v\in\F_q\setminus\{0,1\}$. Thus the roots of the quintic in the nonzero-trace branch are exactly the parameters $v$ for which the point $P(v)=(x(v),y(v))$ lies on the affine line $Y=D-SX.$ Consequently, an ordered pair of distinct roots in one fiber is the same thing as an ordered pair $v \neq w$ whose points $P(v)$ and $P(w)$ lie on the same line of slope $-S$ and intercept $D$.
As we show in Proposition~\ref{prop:secant-parametrization} below, this forces $x(v) \neq x(w)$, and the line is then a unique secant through $P(v)$ and $P(w)$. This converts the higher moments into an incidence problem for secants of the parametrized affine curve $v\longmapsto (x(v),y(v)).$
Solving this two-point interpolation problem gives the functions $K(v,w)$ and $H(v,w)$ in Proposition~\ref{prop:secant-parametrization} and 
the conditions $S=A^2$ with $A\ne0$ and the nonsubfield condition $D=A^2-4B\in C_1$ become, respectively, $K(v,w)\in C_0$ and $H(v,w)\in C_1$.

\begin{proposition}
\label{prop:secant-parametrization}
Let $v,w\in\F_q\setminus\{0,1\}$ with $v\ne w$.  There exists a pair $(S,D)\in\F_q^2$ such that $D=Sx(v)+y(v)=Sx(w)+y(w)$ if and only if $v+w-2\ne0$.  In that case the pair $(S,D)$ is unique and is given by
\begin{equation}
\label{eq:K-secant}
  S=K(v,w):=-\frac{(v-1)^2(w-1)^2(v^2w+2v^2+vw^2+2vw+2w^2)}{v^3w^3(v+w-2)},
\end{equation}
\begin{equation}
\label{eq:H-secant}
  D=H(v,w):=\frac{3v^2w-2v^2+3vw^2-2vw-2w^2}{v^3w^3(v+w-2)}.
\end{equation}
Consequently, admissible pairs $(v,w)$ correspond to ordered pairs of distinct roots in the branch $A\ne0$, with two choices of $A$ for each pair. Their conditions are
\begin{equation}
\label{eq:secant-admissible}
  v,w\in\F_q\setminus\{0,1\},\qquad v\ne w,\qquad v+w-2\ne0,\qquad K(v,w)\in C_0,\qquad H(v,w)\in C_1.
\end{equation}
For each admissible pair $(v,w)$, the corresponding parameters are $A=\pm\sqrt{K(v,w)}$ and $B=(K(v,w)-H(v,w))/4$.
\end{proposition}

\begin{proof}
Suppose first that the same pair $(S,D)$ works for both $v$ and $w$.  Then
$D=Sx(v)+y(v)=Sx(w)+y(w),$
and therefore
$S(x(v)-x(w))=y(w)-y(v).$
A direct calculation gives
$x(v)-x(w)=\frac{(w-v)(v+w-2)}{(v-1)^2(w-1)^2}.$
Thus, if $v\ne w$ and $v+w-2=0$, then $x(v)=x(w)$.  A solution could exist only if also $y(v)=y(w)$.  But for $w=2-v$ one has
$y(v)-y(w)=\frac{16(v-1)}{v^3(v-2)^3},$
which is nonzero for the allowed values $v\ne0,1,2$.  Hence $v+w-2\ne0$ is necessary.

Assume now that $v+w-2\ne0$.  Then $x(v)-x(w)\ne0$, and the preceding linear equation determines $S$ uniquely as
$S=\frac{y(w)-y(v)}{x(v)-x(w)}.$
Substituting the displayed formulas for $x$ and $y$ and simplifying gives exactly \eqref{eq:K-secant}.  Once $S$ is known, the common intercept is forced by
$D=Sx(v)+y(v),$
and simplification gives \eqref{eq:H-secant}.  Therefore the pair $(S,D)$ is unique.

It remains to translate this interpolation statement back to the original parameters.  The condition $S\in C_0$ is exactly the condition that $S=A^2$ for two nonzero values $A=\pm\sqrt S$.  The condition $D\in C_1$ is exactly the condition that $T^2+AT+B$ be irreducible over $\F_q$, where $B=(S-D)/4$.  Thus every admissible ordered pair $(v,w)$ gives the two coefficient pairs
$(A,B)=\left(\pm\sqrt{K(v,w)},\frac{K(v,w)-H(v,w)}4\right),$
and conversely every ordered pair of distinct roots in the nonzero-trace branch gives such an admissible pair $(v,w)$.
\end{proof}

For character-sum manipulations it is useful to write the same count with indicator functions.
If $\mathbf 1_{C_0}$ and $\mathbf 1_{C_1}$ denote the indicators of nonzero squares and nonsquares in $\F_q$, respectively, then Proposition~\ref{prop:secant-parametrization} gives the exact off-diagonal second falling-factorial moment.  The factor $2$ appears because the secant parameters determine $S=A^2$ and $D$, hence two values of $A$, namely $A=\pm\sqrt S$.  For a fixed coefficient pair $(A,B)$ with $n=N(A,B)$ roots in the nonzero-trace branch, the ordered pairs $(v,w)$ with $v\ne w$ contribute exactly $n(n-1)$ terms.  Thus the following sum is precisely the sum of $N(A,B)(N(A,B)-1)$ over all nonzero-trace coefficient pairs, giving
\begin{equation}
\label{eq:F2-nonzero-exact}
  \mathfrak F_2^{\ne0}
  =2\sum_{\substack{v,w\in\F_q\setminus\{0,1\}\\ v\ne w,\ v+w-2\ne0}}
    \mathbf 1_{C_0}(K(v,w))\mathbf 1_{C_1}(H(v,w)).
\end{equation}

The above identity handles ordered pairs of distinct roots. To compute the higher falling factorial moments $\mathfrak F_r$ for $r \geq 3$, we need to describe the remaining roots in the fiber determined by such a pair. We show in Proposition~\ref{prop:residual-cubic} that they are the roots of the following cubic. For an admissible ordered pair $(v,w)$, define $\Phi_{v,w}(Z)\in\F_q[Z]$ by
\begin{align}
\label{eq:residual-cubic}
\Phi_{v,w}(Z)={}&(3v^2w-2v^2+3vw^2-2vw-2w^2)Z^3\notag\\
&+(3v^3w-2v^3+6v^2w^2-10v^2w+4v^2+3vw^3-10vw^2+4vw-2w^3+4w^2)Z^2\notag\\
&+(3v^3w^2-2v^3w+3v^2w^3-10v^2w^2+4v^2w-2vw^3+4vw^2)Z\notag\\
&-2v^3w^2-2v^2w^3+4v^2w^2 .
\end{align}

\begin{proposition}
\label{prop:residual-cubic}
Let $(v,w)$ satisfy the admissibility conditions \eqref{eq:secant-admissible}, and put $S=K(v,w)$, $D=H(v,w)$.  Then
\begin{equation}
\label{eq:residual-factorization}
  v^3w^3(v+w-2)\bigl(DZ^3(Z-1)^2-(S+1)Z^3+3Z-2\bigr)=(Z-v)(Z-w)\Phi_{v,w}(Z).
\end{equation}
Consequently, the remaining roots in the same fiber are exactly the roots of $\Phi_{v,w}$ different from $v$ and $w$.
\end{proposition}

\begin{proof}
For $Z\in\F_q\setminus\{0,1\}$ the condition $D=S\,x(Z)+y(Z)$ is equivalent to
\[
  Q_{S,D}(Z):=DZ^3(Z-1)^2-(S+1)Z^3+3Z-2=0 .
\]
Since $v$ and $w$ satisfy \eqref{eq:D-line-form} with the same pair $(S,D)$, both are roots of  $Q_{S,D}(Z)$.  Since $v \neq w$, the product $(Z-v)(Z-w)$ divides $Q_{S,D}(Z)$. Substituting \eqref{eq:K-secant} and \eqref{eq:H-secant} for $S$
and $D$, clearing the common denominator $v^3w^3(v+w-2)$, which is nonzero by
admissibility, and dividing by $(Z-v)(Z-w)$ gives exactly the cubic
\eqref{eq:residual-cubic}. This proves \eqref{eq:residual-factorization}.

Finally, $Q_{S,D}(0)=-2$ and $Q_{S,D}(1)=-S$ are nonzero. Hence, the roots in the fiber $(S,D)$ are exactly the roots of $Q_{S,D}$ in $\F_q$, which by \eqref{eq:residual-factorization} are $v,w$ and roots of $\Phi_{v,w}(Z) \in \F_q$.
\end{proof}

Let $\mathcal E$ denote the set of ordered pairs satisfying \eqref{eq:secant-admissible}. For each $(v,w)\in\mathcal E$, put
\begin{equation*}
  \rho(v,w)=\#\{z\in\F_q\setminus\{0,1,v,w\}:\Phi_{v,w}(z)=0\}.
\end{equation*}
This number is the number of further roots in the same fiber after the ordered pair $(v,w)$ has been fixed.  Hence, if a nonzero-trace fiber has $n$ roots, then every one of its $n(n-1)$ ordered pairs contributes $n-2$ choices to the third falling-factorial count and $(n-2)(n-3)$ choices to the fourth falling-factorial count.  Summing this over all ordered admissible pairs, and again multiplying by $2$ for the two signs of $A$, gives the exact nonzero-trace contributions
\begin{equation}
\label{eq:F3F4-nonzero-exact}
  \mathfrak F_3^{\ne0}=2\sum_{(v,w)\in\mathcal E}\rho(v,w),
  \qquad
  \mathfrak F_4^{\ne0}=2\sum_{(v,w)\in\mathcal E}\rho(v,w)(\rho(v,w)-1),
\end{equation}

To obtain the full moments, we add the exceptional trace-zero branch.  Let $\mathcal B_0=\{B\in\F_q:-4B\in C_1\}$.  For $B\in\mathcal B_0$, set
$N_0(B)=\#\{v\in\F_q\setminus\{1\}:4Bv^3+v+2=0\}+1+\eta(-4B-3),$
as in Proposition~\ref{prop:A-zero}, and define
$\mathfrak F_r^0=\sum_{B\in\mathcal B_0}N_0(B)(N_0(B)-1)\cdots(N_0(B)-r+1)\ (r\ge1).$
Combining the trace-zero and nonzero-trace branches gives $\mathfrak F_r=\mathfrak F_r^0+\mathfrak F_r^{\ne0}$.  The ordinary moments used in the Vandermonde system are obtained from the falling-factorial moments by
\begin{align}
\label{eq:moment-from-falling}
M_1&=\mathfrak F_1,\notag\\
M_2&=\mathfrak F_1+\mathfrak F_2,\notag\\
M_3&=\mathfrak F_1+3\mathfrak F_2+\mathfrak F_3,\notag\\
M_4&=\mathfrak F_1+7\mathfrak F_2+6\mathfrak F_3+\mathfrak F_4.
\end{align}
Equations \eqref{eq:F2-nonzero-exact}, \eqref{eq:F3F4-nonzero-exact}, and \eqref{eq:moment-from-falling} are exact finite-field formulas.

The moments can also be interpreted as counts of tuples satisfying the original coordinate system.
For $r\ge1$, recall that
\begin{equation*}
  M_r=\sum_{(A,B)\in\mathcal D'}N(A,B)^r.
\end{equation*}
Then $M_r$ is the number of tuples
  $(A,B,u_1,v_1,\ldots,u_r,v_r)\in\F_q^{2+2r}$
such that $(A,B)\in\mathcal D'$ and each pair $(u_j,v_j)$ satisfies \eqref{eq:uv-system}.  For the following identity, extend $N(A,B)$ to all coefficient pairs by the number of solutions of \eqref{eq:uv-system}. Square-discriminant pairs have zero weight. Then
\begin{equation*}
  M_r=\frac12\sum_{\substack{A,B\in\F_q\\ A^2-4B \ne0}}
  \bigl(1-\eta(A^2-4B)\bigr)N(A,B)^r.
\end{equation*}
This character filter expresses the moments as point counts for the coordinate system~\eqref{eq:uv-system}.

\begin{theorem}
\label{thm:spectrum-from-moments}
Let $q\equiv1\pmod6$. Put $c=m_4$, which is zero for $q\equiv1\pmod{12}$ and is given by Theorem~\textup{\ref{thm:N4-zero2}} for $q\equiv7\pmod{12}$. The remaining nonsubfield frequencies $m_0,m_1,m_2,m_3,m_5$ are the unique solution to the linear system
\begin{equation}
\label{eq:moment-vandermonde}
\small
  \begin{pmatrix}
    1 & 1 & 1 & 1 & 1 \\
    0 & 1 & 2 & 3 & 5 \\
    0 & 1 & 4 & 9 & 25 \\
    0 & 1 & 8 & 27 & 125 \\
    0 & 1 & 16 & 81 & 625
  \end{pmatrix}
  \begin{pmatrix} m_0 \\ m_1 \\ m_2 \\ m_3 \\ m_5 \end{pmatrix}
  =
  \begin{pmatrix} M_0-c \\ M_1-4c \\ M_2-16c \\ M_3-64c \\ M_4-256c \end{pmatrix},
\end{equation}
where $M_0=(q^2-q)/2$ and $M_1,M_2,M_3,M_4$ are given by
\eqref{eq:moment-from-falling}.  The coefficient matrix is the Vandermonde matrix
on $\{0,1,2,3,5\}$ and is nonsingular.  Therefore the nonsubfield distribution is
determined from $M_0,\ldots,M_4$ and $c$ once the nonzero-trace sums in \eqref{eq:F2-nonzero-exact} and \eqref{eq:F3F4-nonzero-exact}, together with $\mathfrak F_2^0,\mathfrak F_3^0,\mathfrak F_4^0$, are evaluated. The first moment is already evaluated in Theorem~\textup{\ref{thm:first-moment-exact}}.
\end{theorem}

\begin{proof}
Recall that $M_0=(q^2-q)/2$ and for $r\ge1$, the definition of the ordinary moments is $M_r=\sum_{(A,B)\in\mathcal D'}N(A,B)^r.$
The quintic reduction and Proposition~\ref{prop:A-zero} show that $N(A,B)\le5$, therefore $M_r=\sum_{i=0}^{5} i^r m_i$ for every $r\ge 1$. Theorems~\ref{thm:N4-zero} and~\ref{thm:N4-zero2} determine $c=m_4$. Removing the size-four contribution gives $M_r-4^r c=\sum_{i\in\{0,1,2,3,5\}}i^r m_i$ for $r=0,1,2,3,4$, with the zeroth equation interpreted as a sum of frequencies.  These five
relations are exactly \eqref{eq:moment-vandermonde}.

The determinant of the coefficient matrix is the Vandermonde product
$\prod_{0\le a<b,\ a,b\in\{0,1,2,3,5\}}(b-a),$
viewed as an integer.  It is nonzero, so the coefficient matrix is invertible
over $\mathbb{Q}$.  The moment system is a system of integer linear equations,
not a linear system over the finite field $\F_q$, and therefore no reduction modulo
the characteristic is involved.  Thus the system has a unique solution.  Finally,
\eqref{eq:moment-from-falling} expresses the ordinary moments in terms of the
falling-factorial moments, while \eqref{eq:F2-nonzero-exact} and
\eqref{eq:F3F4-nonzero-exact} give those falling-factorial moments exactly.
\end{proof}

\begin{corollary}
\label{cor:explicit-frequencies}
With $c=m_4$ as above, the frequencies can be recovered explicitly by
\begin{align*}
m_5&=(\mathfrak F_4-24c)/120,\\
m_3&=(\mathfrak F_3-24c-60m_5)/6,\\
m_2&=(\mathfrak F_2-6m_3-12c-20m_5)/2,\\
m_1&=M_1-2m_2-3m_3-4c-5m_5,\\
m_0&=M_0-m_1-m_2-m_3-c-m_5.
\end{align*}
In particular, the fifth ordinary moment is unnecessary in both congruence classes.
\end{corollary}

\begin{proof}
For fiber sizes at most five, $\mathfrak F_4=24m_4+120m_5$, $\mathfrak F_3=6m_3+24m_4+60m_5$, and $\mathfrak F_2=2m_2+6m_3+12m_4+20m_5$. Solve these equations successively, then use the first and zeroth ordinary moments. All divisions take place in $\mathbb Q$.
\end{proof}

Tables~\ref{tab:q1mod12} and~\ref{tab:q7mod12} record the fiber distributions $m_i$ for small prime powers, computed by direct enumeration.  The domain size
$\#\mathcal D'=(q^2-q)/2$ was verified in all cases. The Python script \texttt{boomerang\_tables.py} in Appendix~\ref{app:python-tables} reproduces these tables and Tables~\ref{tab:n-q1mod12} and~\ref{tab:n-q7mod12} by enumerating the roots of~\eqref{eq:main-equation}. It independently checks the quintic and trace-zero reductions, the subfield distribution, the first-moment formulas, and the size-four characterization. The script uses only the Python standard library.

\begin{table}[htbp]
\centering
\small
\renewcommand{\arraystretch}{1.2}
\begin{tabular}{r|rrrrrrrrrrrr}
$q$ & $m_0$ & $m_1$ & $m_2$ & $m_3$ & $m_4$ & $m_5$ & $n_0^{(q)}$ & $n_1^{(q)}$ & $n_2^{(q)}$ & $n_3^{(q)}$ & $n_4^{(q)}$ & $n_5^{(q)}$ \\
\hline
$13$ & $25$  & $42$   & $5$   & $5$   & $0$ & $1$  & $4$  & $0$ & $2$ & $6$  & $0$ & $0$ \\
$25$ & $ 116$ & $ 108$  & $ 48$ & $28 $  & $0 $ & $ 0$  & $8$ & $0$ & $2$ & $ 14$ & $0$ & $0$ \\
$37$ & $245$ & $258$  & $109$ & $49$  & $0$ & $5$  & $12$ & $0$ & $2$ & $22$ & $0$ & $0$ \\
$61$ & $690$ & $654$  & $336$ & $138$ & $0$ & $12$ & $20$ & $0$ & $2$ & $38$ & $0$ & $0$ \\
$73$ & $952$ & $1032$ & $420$ & $200$ & $0$ & $24$ & $24$ & $0$ & $2$ & $46$ & $0$ & $0$ \\
\end{tabular}
\caption{Values of $m_i$ and $n_i^{(q)}$ for $q \equiv 1 \pmod{12}$.}
\label{tab:q1mod12}
\end{table}

\begin{table}[htbp]
\centering
\small
\renewcommand{\arraystretch}{1.2}
\begin{tabular}{r|rrrrrrrrrrrr}
$q$ & $m_0$ & $m_1$ & $m_2$ & $m_3$ & $m_4$ & $m_5$ & $n_0^{(q)}$ & $n_1^{(q)}$ & $n_2^{(q)}$ & $n_3^{(q)}$ & $n_4^{(q)}$ & $n_5^{(q)}$ \\
\hline
$7$  & $8$    & $9$    & $4$   & $0$   & $0$  & $0$  & $2$  & $0$ & $0$ & $4$  & $0$ & $0$ \\
$19$ & $65$   & $65$   & $28$  & $12$  & $1$  & $0$  & $6$  & $0$ & $0$ & $12$ & $0$ & $0$ \\
$31$ & $167$  & $188$  & $82$  & $18$  & $7$  & $3$  & $10$ & $0$ & $0$ & $20$ & $0$ & $0$ \\
$43$ & $329$  & $354$  & $151$ & $57$  & $6$  & $6$  & $14$ & $0$ & $0$ & $28$ & $0$ & $0$ \\
$79$ & $1111$ & $1212$ & $514$ & $202$ & $15$ & $27$ & $26$ & $0$ & $0$ & $52$ & $0$ & $0$ \\
\end{tabular}
\caption{Values of $m_i$ and $n_i^{(q)}$ for $q \equiv 7 \pmod{12}$.}
\label{tab:q7mod12}
\end{table}

Note that $m_4=0$ holds in Table~\ref{tab:q1mod12},
consistent with Theorem~\ref{thm:N4-zero}.  The entries in Table~\ref{tab:q7mod12} are consistent with the exact characterization $N=4$ in Theorem~\ref{thm:N4-zero2}.

\begin{remark}
The fixed-trace bound in Theorem~\textup{\ref{thm:incidence-bound}} concerns first-moment incidences. It does not determine the asymptotic behavior of the individual frequencies $m_i$. The exact first moment is given by Theorem~\textup{\ref{thm:first-moment-exact}}, while evaluations of the higher-moment sums remain open.
\end{remark}

\section{Consequences for the nonsubfield distribution}
\label{sec:nonsubfield-consequences}

Each coefficient pair $(A,B)\in\mathcal D'$ determines the irreducible quadratic $T^2+AT+B$ and its two conjugate roots in $\F_{q^2}\setminus\F_q$. Both roots have the same value of $N(A,B)$, so the coefficient-pair frequencies give the nonsubfield contribution with a factor of two.

\begin{proposition}
\label{prop:ns-spectrum-exact}
For $0\le i\le5$, the nonsubfield contribution to the normalized spectrum is
\begin{equation}
\label{eq:nis-exact}
  n_i^{\mathrm{ns}}
  =2\cdot\#\{(A,B)\in\mathcal D':N(A,B)=i\}.
\end{equation}
Consequently, with $n_i^{(q)}$ as in \Cref{thm:subfield-theorem}, the full normalized spectrum is
\begin{equation}
\label{eq:full-normalized-spectrum}
  n_i=n_i^{(q)}+n_i^{\mathrm{ns}},
  \qquad
  \nu_i=(q^2-1)n_i.
\end{equation}
\end{proposition}

\begin{proof}
The map
\[
  b\longmapsto (A,B)=(-(b+b^q),b^{q+1})
\]
identifies the unordered conjugate pair $\{b,b^q\}$ with the coefficients of the irreducible quadratic $T^2+AT+B$ over $\F_q$.  Each pair $(A,B)\in\mathcal D'$ therefore corresponds to exactly two elements of $\F_{q^2}\setminus\F_q$, namely the two roots of that quadratic.  The number of roots of \eqref{eq:main-equation} attached to either conjugate is $N(A,B)$, because the system \eqref{eq:uv-system} was obtained by expanding \eqref{eq:main-equation} in the basis $\{1,b\}$ and separating the $\F_q$-coordinates.  This proves \eqref{eq:nis-exact}.  The identity \eqref{eq:full-normalized-spectrum} follows by adding the already computed subfield contribution and then applying the power-map normalization in Lemma~\ref{lem:power-normalization}.
\end{proof}

For small prime powers $q$, applying the relation
$\nu_i=(q^2-1)(n_i^{(q)}+n_i^{\mathrm{ns}})=(q^2-1)(n_i^{(q)}+2m_i)$ to the values in Tables~\ref{tab:q1mod12} and~\ref{tab:q7mod12} yields the complete boomerang spectrum of $x^{q+2}$ over $\F_{q^2}$, presented in Tables~\ref{tab:n-q1mod12} and~\ref{tab:n-q7mod12}.

\begin{table}[htbp]
\centering
\small
\renewcommand{\arraystretch}{1.2}
\begin{tabular}{crrrrrr}
$q$  & $\nu_0$ & $\nu_1$ & $\nu_2$ & $\nu_3$ & $\nu_4$ & $\nu_5$ \\ \hline
$13$   & $9072$     & $14112$    & $2016$    & $2688$    & $0$ & $336$    \\
$25$ & $ 149760$ & $ 134784$  & $ 61152$ & $43680 $  & $0 $ & $ 0$   \\
$37$  & $686736$   & $705888$   & $300960$  & $164160$  & $0$ & $13680$  \\
$61$  & $5208000$  & $4865760$  & $2507280$ & $1168080$ & $0$ & $89280$  \\
$73$  & $10272384$ & $10996992$ & $4486176$ & $2376288$ & $0$ & $255744$ \\
\end{tabular}
\caption{Boomerang spectrum of $x^{q+2}$ for $q \equiv 1 \pmod{12}$.}
\label{tab:n-q1mod12}
\end{table}

\begin{table}[htbp]
\centering
\small
\renewcommand{\arraystretch}{1.2}
\begin{tabular}{crrrrrr}
$q$  & $\nu_0$ & $\nu_1$ & $\nu_2$ & $\nu_3$ & $\nu_4$ & $\nu_5$ \\ \hline
$7$    & $864$      & $864$      & $384$     & $192$     & $0$      & $0$      \\
$19$   & $48960$    & $46800$    & $20160$   & $12960$   & $720$    & $0$      \\
$31$   & $330240$   & $360960$   & $157440$  & $53760$   & $13440$  & $5760$   \\
$43$ & $1241856$  & $1308384$  & $558096$  & $262416$  & $22176$  & $22176$  \\
$79$  & $14027520$ & $15125760$ & $6414720$ & $2845440$ & $187200$ & $336960$ \\
\end{tabular}
\caption{Boomerang spectrum of $x^{q+2}$ for $q \equiv 7 \pmod{12}$.}
\label{tab:n-q7mod12}
\end{table}

We now assemble the preceding analysis and prove the main boomerang-spectrum theorem.  The subfield and nonsubfield cases are disjoint and exhaustive, and the power-map normalization then lifts the normalized counts to the full spectrum.

\begin{theorem}
\label{thm:main_boomerang}
Let $q=p^m$ be an odd prime power with $q\equiv1\pmod6$, and let $F(x)=x^{q+2}$ on $\F_{q^2}$.  For $b\in\F_{q^2}^*$, put $N_F(b)=\#\{X\in\F_{q^2}:4X^{q+2}+X^q+2X+2b=0\}$, and set $n_i=\#\{b\in\F_{q^2}^*:N_F(b)=i\}$ and $\nu_i=\#\{(a,b)\in(\F_{q^2}^*)^2:\beta_F(a,b)=i\}$.  Every entry is at most five, and the boomerang spectrum of $F$ is $\BS(F)=(\nu_i)_{0\le i\le\BU(F)}$, where $\nu_i=(q^2-1)n_i$ and
\begin{equation}
\label{eq:main-spectrum-decomposition}
  n_i=n_i^{(q)}+2m_i.
\end{equation}
Here $n_i^{(q)}$ is the subfield contribution
\[
\begin{array}{c|cccccc}
 & n_0^{(q)}&n_1^{(q)}&n_2^{(q)}&n_3^{(q)}&n_4^{(q)}&n_5^{(q)}\\ \hline
q\equiv1\pmod{12} &(q-1)/3&0&2&(2q-8)/3&0&0\\[1mm]
q\equiv7\pmod{12} &(q-1)/3&0&0&2(q-1)/3&0&0,
\end{array}
\]
and $m_i$ is the nonsubfield trace--norm contribution defined as follows.  Let $\mathcal D'=\{(A,B)\in\F_q^2:A^2-4B\in C_1\}$, where $C_1$ is the set of nonsquares of $\F_q^*$.  For $(A,B)\in\mathcal D'$, define
\begin{equation*}
N(A,B)=
\begin{cases}
\#\{v\in\F_q\setminus\{0,1\}:Q_{A,B}(v)=0\},& A\ne0,\\[1mm]
\#\{v\in\F_q\setminus\{1\}:4Bv^3+v+2=0\}+1+\eta(-4B-3),& A=0,
\end{cases}
\end{equation*}
where
\[
  Q_{A,B}(V)=(A^2-4B)V^5-2(A^2-4B)V^4-(4B+1)V^3+3V-2.
\]
Then $m_i=\#\{(A,B)\in\mathcal D':N(A,B)=i\}$.  Equivalently, each pair $(A,B)\in\mathcal D'$ corresponds to the two conjugate nonsubfield parameters $b,b^q$, where $b$ is a root of $T^2+AT+B$, and both contribute the same value $N(A,B)$ to the normalized spectrum.
\end{theorem}

\begin{proof}
The equation defining $N_F(b)$ is equivalent to the boomerang entry $\beta_F(1,b)$ by Proposition~\textup{\ref{prop:working-reduction}}.  Lemma~\textup{\ref{lem:power-normalization}} then gives $\nu_i=(q^2-1)n_i$.  The contribution from $b\in\F_q^*$ is Theorem~\textup{\ref{thm:subfield-theorem}}, and the contribution from $b\in\F_{q^2}\setminus\F_q$ is Proposition~\textup{\ref{prop:ns-spectrum-exact}}.  Combining these two disjoint cases gives \eqref{eq:main-spectrum-decomposition}. The subfield bound is three, the nonzero-trace bound is five by the quintic degree, and the trace-zero bound is five by Proposition~\ref{prop:A-zero}. Thus there are no larger entries.
\end{proof}

\section{Conclusion}
\label{sec:conclusion}

We have determined the complete contribution of nonzero subfield parameters to the boomerang spectrum of $x^{q+2}$ over $\F_{q^2}$ for $q\equiv1\pmod6$. For nonsubfield parameters, the trace--norm reduction gives a quintic root count when the trace is nonzero and a separate cubic and quadratic count when the trace is zero. The repeated-root calculation excludes the entry four when $q\equiv1\pmod{12}$ and characterizes it through a residual cubic when $q\equiv7\pmod{12}$.

The exact first moment is $M_1=(q^2-2q+\eta(-1))/2$. The fixed-trace incidence formula also gives a Weil bound and exact aggregate character and zero counts. Secant interpolation expresses the higher falling-factorial moments as explicit finite sums. Once the size-four frequency is known, the moments through order four recover every remaining frequency by Corollary~\ref{cor:explicit-frequencies}.

A closed-form nonsubfield distribution still requires the evaluation of the second, third, and fourth falling-factorial sums, including their trace-zero contributions. When $q\equiv7\pmod{12}$, it also requires the residual-cubic splitting count that determines the size-four frequency. The formulas proved here specify these remaining finite-field counting problems and recover the complete spectrum for each fixed $q$ by enumeration.


\noindent
\subsection*{\bf Ethics and consent to participate:} Not applicable.

\noindent
\subsection*{\bf Consent to publish:}  Not applicable.

\noindent
\subsection*{\bf Availability of data and materials:} The datasets used during the current study are available from the corresponding author upon reasonable request.

\noindent
\subsection*{\bf Competing interests:} The authors declare that they have no competing interests.

\noindent
\subsection*{\bf Funding:} The authors received no financial support for the research, authorship, or publication of this article.

\noindent
\subsection*{\bf Authors' contributions:} All authors contributed equally, read and approved the final manuscript.

\noindent
\subsection*{\bf Acknowledgements:} Not applicable.

\appendix
\section{Python code for the numerical tables}
\label{app:python-tables}

The following Python script (generated by OpenAI - Sol 6.1 to double check the output of our prior SageMath code -- we note that the AI agent code is more efficient -- data matches perfectly) reproduces Tables~\ref{tab:q1mod12}, \ref{tab:q7mod12}, \ref{tab:n-q1mod12}, and~\ref{tab:n-q7mod12} using Python~3.9 or later and only the standard library. For each $X\in\F_{q^2}$, it computes the unique output parameter  $b=-\frac{4X^{q+2}+X^q+2X}{2}$
and counts the preimages of every $b$. It then separates the nonzero subfield parameters from the nonsubfield parameters and groups the latter into Frobenius conjugate pairs to obtain $n_i^{(q)}$ and $m_i$. The full spectrum follows from $\nu_i=(q^2-1)(n_i^{(q)}+2m_i)$. The script stores no numerical table entries and checks the computed counts against the quintic and trace-zero reductions, the subfield distribution, the first-moment formulas, and the size-four characterization.

The implemented ground fields are the prime fields with $q\equiv1\pmod6$ and $\F_{25}=\F_5[t]/(t^2-2)$, which cover all parameters in the four tables. For each ground field, the quadratic extension is represented as $\F_q[s]/(s^2-d)$, where $d$ is a nonsquare in $\F_q$. All arithmetic is exact. The optional \texttt{--check-bct} flag also verifies the normalized counts directly from the coupled system~\eqref{eq:boomerang-system}, independently of the single-equation reduction.

Save the listing as \texttt{boomerang\_tables.py} and run \texttt{python3 boomerang\_tables.py} to print all four tables as comma-separated rows. Adding \texttt{--latex} prints LaTeX tabular environments. 
These computations verify the numerical examples and are not used in the proofs of the general results.

\begin{lstlisting}[style=pythonappendix]
#!/usr/bin/env python3
"""Reproduce the four numerical tables in "On the boomerang spectrum of the APN power function x^{q+2} over F_{q^2}", by K. Garg, C. Riera, P. Stanica.
Python 3.9 or later. No third-party packages are required.

    python3 boomerang_tables.py
    python3 boomerang_tables.py --latex > boomerang_tables.tex
    python3 boomerang_tables.py --check-tex 'BoomerangSpectrum(2).tex'
    python3 boomerang_tables.py --q 7 13 --check-bct

The default q values are exactly those in the paper. Other primes q = 1 mod 6
are also supported, as is q = 25. No other nonprime q is implemented.

We enumerate X in F_(q^2), compute  b = -(4 X^(q+2) + X^q + 2 X)/2,
and tally N_F(b). Thus the primary computation does not use the quintic.
We then check every nonsubfield fiber against the quintic or trace-zero formula,
and check the subfield distribution, first moments, and size-four formula.
The optional --check-bct independently counts the original coupled system at
a = 1. The optional --check-tex compares computed rows with all four tables
in the supplied manuscript. No published table entries are stored in this file.

All field arithmetic is exact. F_25 is F_5[t]/(t^2-2), with a+bt encoded as
a+5b. We construct F_(q^2) as F_q[s]/(s^2-d), where d is the first nonsquare
in this encoding. Elements of this quadratic extension are pairs (a,b).
"""

import argparse
import math
import re
import sys
from collections import Counter, defaultdict
from pathlib import Path

DEFAULT_Q = (7, 13, 19, 25, 31, 37, 43, 61, 73, 79)

def require(condition, message):
    """Keep verification enabled even when Python is run with -O."""
    if not condition:
        raise ValueError(message)

class Field:
    """Prime fields and the explicit quadratic field F_25."""

    def __init__(self, q):
        prime = q >= 2 and all(q % k for k in range(2, math.isqrt(q) + 1))
        require(q == 25 or prime, "Only prime q and q = 25 are supported.")
        require(q % 6 == 1, "The parameter q must satisfy q = 1 mod 6.")
        self.q = q
        self.p = 5 if q == 25 else q

    def scalar(self, n):
        return n % self.p

    def add(self, a, b):
        if self.q == 25:
            return (a % 5 + b % 5) % 5 + 5 * ((a // 5 + b // 5) % 5)
        return (a + b) % self.q

    def neg(self, a):
        if self.q == 25:
            return (-(a % 5)) % 5 + 5 * ((-(a // 5)) % 5)
        return (-a) % self.q

    def sub(self, a, b):
        return self.add(a, self.neg(b))

    def mul(self, a, b):
        if self.q == 25:
            a0, a1 = a % 5, a // 5
            b0, b1 = b % 5, b // 5
            return (a0 * b0 + 2 * a1 * b1) % 5 + 5 * ((a0 * b1 + a1 * b0) % 5)
        return a * b % self.q

    def power(self, a, n):
        result = 1
        while n:
            if n & 1:
                result = self.mul(result, a)
            a = self.mul(a, a)
            n //= 2
        return result

    def div(self, a, b):
        require(b != 0, "Division by zero in F_q.")
        return self.mul(a, self.power(b, self.q - 2))

    def eta(self, a):
        return 0 if a == 0 else (1 if self.power(a, (self.q - 1) // 2) == 1 else -1)

    def polynomial(self, coefficients, x):
        """Evaluate coefficients in descending degree by Horner's rule."""
        result = 0
        for a in coefficients:
            result = self.add(self.mul(result, x), a)
        return result

class QuadraticExtension:
    def __init__(self, field):
        self.f = field
        self.d = next(a for a in range(1, field.q) if field.eta(a) == -1)

    def add(self, x, y):
        return self.f.add(x[0], y[0]), self.f.add(x[1], y[1])

    def sub(self, x, y):
        return self.f.sub(x[0], y[0]), self.f.sub(x[1], y[1])

    def scale(self, c, x):
        return self.f.mul(c, x[0]), self.f.mul(c, x[1])

    def mul(self, x, y):
        f = self.f
        return (f.add(f.mul(x[0], y[0]), f.mul(self.d, f.mul(x[1], y[1]))),
                f.add(f.mul(x[0], y[1]), f.mul(x[1], y[0])))

    def power(self, x, n):
        result = (1, 0)
        while n:
            if n & 1:
                result = self.mul(result, x)
            x = self.mul(x, x)
            n //= 2
        return result

def reduction_count(f, A, B):
    """The paper's quintic count, with the separate A = 0 formula."""
    c, add, sub, mul = f.scalar, f.add, f.sub, f.mul
    if A:
        D = sub(mul(A, A), mul(c(4), B))
        coefficients = [D, mul(c(-2), D), f.neg(add(mul(c(4), B), 1)), 0, c(3), c(-2)]
        return sum(f.polynomial(coefficients, v) == 0 for v in range(f.q) if v not in (0, 1))
    cubic = [mul(c(4), B), 0, 1, c(2)]
    return (sum(f.polynomial(cubic, v) == 0 for v in range(f.q) if v != 1)
            + 1 + f.eta(sub(mul(c(-4), B), c(3))))

def compute(q, check_bct=False):
    f = Field(q)
    ext = QuadraticExtension(f)
    c, sub, mul = f.scalar, f.sub, f.mul
    elements = [(a, b) for a in range(q) for b in range(q)]
    powers = {x: ext.power(x, q + 2) for x in elements}
    counts = Counter()
    minus_half = f.div(c(-1), c(2))
    for x in elements:
        value = ext.add(ext.scale(c(4), powers[x]), ext.power(x, q))
        value = ext.add(value, ext.scale(c(2), x))
        counts[ext.scale(minus_half, value)] += 1
    require(counts[(0, 0)] == 3, f"q={q}: incorrect zero-output count.")

    # Extract subfield frequencies and unordered Frobenius-conjugate pairs.
    ns_pairs = {}
    subfield = Counter(counts[(a, 0)] for a in range(1, q))
    for a, b in elements:
        if not b:
            continue
        A = mul(c(-2), a)
        B = sub(mul(a, a), mul(ext.d, mul(b, b)))
        n = counts[(a, b)]
        if (A, B) in ns_pairs:
            require(ns_pairs[A, B] == n, f"q={q}: conjugate counts differ.")
        else:
            ns_pairs[A, B] = n
            require(reduction_count(f, A, B) == n, f"q={q}: reduction failed at {(A, B)}.")

    m = Counter(ns_pairs.values())
    eps = f.eta(c(-1))
    subfield_formula = ([(q - 1) // 3, 0, 2, (2 * q - 8) // 3, 0, 0]
                        if eps == 1 else [(q - 1) // 3, 0, 0, 2 * (q - 1) // 3, 0, 0])
    require(max(counts.values()) <= 5, f"q={q}: root bound failed.")
    require(len(ns_pairs) == q * (q - 1) // 2, f"q={q}: domain cardinality failed.")
    require([subfield[i] for i in range(6)] == subfield_formula, f"q={q}: subfield formula failed.")
    require(sum(i * n for i, n in m.items()) == (q * q - 2 * q + eps) // 2,
            f"q={q}: first-moment formula failed.")
    require(sum(n for (A, B), n in ns_pairs.items() if A == 0) == q - (3 - eps) // 2,
            f"q={q}: trace-zero first moment failed.")

    if eps == 1:
        require(m[4] == 0, f"q={q}: exclusion of four failed.")
    else:
        splitting_count = 0
        for v in range(q):
            if v in (0, 1, c(-3)) or f.eta(mul(sub(v, 1), f.add(v, c(3)))) != -1:
                continue
            cubic = [c(3), mul(c(6), sub(v, 1)),
                     sub(mul(c(3), mul(v, v)), mul(c(4), v)), mul(c(-2), mul(v, v))]
            splitting_count += sum(f.polynomial(cubic, z) == 0 for z in range(q)) == 3
        require(m[4] == (1 + f.eta(c(5))) // 2 + 2 * splitting_count,
                f"q={q}: size-four formula failed.")

    if check_bct:
        # Equal derivatives group exactly those (x,y) satisfying both equations
        # F(x)-F(y)=b and F(x+1)-F(y+1)=b. Count all such ordered pairs.
        groups = defaultdict(list)
        for x in elements:
            groups[ext.sub(powers[ext.add(x, (1, 0))], powers[x])].append(x)
        direct = Counter()
        for group in groups.values():
            for x in group:
                for y in group:
                    output = ext.sub(powers[x], powers[y])
                    if output != (0, 0):
                        direct[output] += 1
        require(all(direct[b] == counts[b] for b in elements if b != (0, 0)),
                f"q={q}: coupled-system verification failed.")

    m_row = [m[i] for i in range(6)]
    subfield_row = [subfield[i] for i in range(6)]
    nu_row = [(q * q - 1) * (2 * m_row[i] + subfield_row[i]) for i in range(6)]
    require(sum(nu_row) == (q * q - 1) ** 2, f"q={q}: spectrum total failed.")
    return m_row + subfield_row, nu_row

def check_manuscript(path, results):
    source = Path(path).read_text(encoding="utf-8")
    source = "\n".join(line for line in source.splitlines() if not line.lstrip().startswith("%"))
    tables = re.findall(r"\\begin\{table\}.*?\\end\{table\}", source, re.S)
    for label, residue, index in (("tab:q1mod12", 1, 0), ("tab:q7mod12", 7, 0),
                                  ("tab:n-q1mod12", 1, 1), ("tab:n-q7mod12", 7, 1)):
        selected = [t for t in tables if r"\label{" + label + "}" in t]
        require(len(selected) == 1, f"Could not uniquely locate table {label}.")
        rows = {}
        for match in re.finditer(r"^\s*\$(\d+)\$\s*&([^\n]+)", selected[0], re.M):
            q = int(match.group(1))
            cells = match.group(2).split(r"\\")[0].split("&")
            rows[q] = [int(cell.replace("$", "").strip()) for cell in cells]
        for q, computed in results.items():
            if q % 12 == residue:
                require(rows.get(q) == computed[index], f"Table {label}, q={q}: row mismatch.")


def print_tables(results, latex):
    for index, title, headers in ((0, "Coefficient-pair and subfield frequencies",
                                  [f"m_{i}" for i in range(6)] + [f"n_{i}^(q)" for i in range(6)]),
                                 (1, "Full boomerang spectrum", [f"nu_{i}" for i in range(6)])):
        for residue in (1, 7):
            rows = [(q, result[index]) for q, result in sorted(results.items()) if q % 12 == residue]
            if not rows:
                continue
            print(("% " if latex else "") + f"{title}, q = {residue} mod 12")
            if latex:
                print(r"\begin{tabular}{r|" + "r" * len(headers) + "}")
                symbols = ([f"m_{i}" for i in range(6)] + [f"n_{i}^{{(q)}}" for i in range(6)]
                           if index == 0 else [rf"\nu_{i}" for i in range(6)])
                print("$q$ & " + " & ".join(f"${h}$" for h in symbols) + r" \\ \hline")
                for q, values in rows:
                    print(" & ".join(f"${v}$" for v in [q] + values) + r" \\")
                print(r"\end{tabular}")
            else:
                print("q," + ",".join(headers))
                for q, values in rows:
                    print(",".join(map(str, [q] + values)))
            print()

def main():
    parser = argparse.ArgumentParser(description=__doc__, formatter_class=argparse.RawDescriptionHelpFormatter)
    parser.add_argument("--q", nargs="+", type=int, default=DEFAULT_Q, help="Field parameters (default: all paper rows).")
    parser.add_argument("--latex", action="store_true", help="Print four LaTeX tabular environments.")
    parser.add_argument("--check-bct", action="store_true", help="Also count the original coupled boomerang system.")
    parser.add_argument("--check-tex", type=Path, help="Compare the computed rows with the four manuscript tables.")
    args = parser.parse_args()
    results = {}
    for q in sorted(set(args.q)):
        results[q] = compute(q, args.check_bct)
        print(f"q={q}: verification passed", file=sys.stderr)
    if args.check_tex:
        check_manuscript(args.check_tex, results)
        print("All selected manuscript rows agree.", file=sys.stderr)
    print_tables(results, args.latex)

if __name__ == "__main__":
    main()
\end{lstlisting}

\end{document}